\documentclass[twocolumn,showkeys,showpacs,preprintnumbers,amsmath,amssymb,superscriptaddress]{IEEEtran}

\usepackage{amsmath,amssymb,amsfonts}
\usepackage{graphicx}
\usepackage{subfigure}
\usepackage{url}

\begin{document}

\title{Controlling symmetries and clustered dynamics of complex networks}

\author{L. V. Gambuzza, M. Frasca, \IEEEmembership{Senior, IEEE},  F. Sorrentino, \IEEEmembership{Senior, IEEE}, L. M. Pecora, and S. Boccaletti
\thanks{LVG and MF acknowledge the support of the Italian Ministry for Research and Education through the Research Program PRIN 2017 (Grant 2017CWMF93, project VECTORS). LMP acknowledges the support of the Office of Naval Research through the Naval Research Laboratory's Basic Research Program.}
\thanks{L. V. Gambuzza is with Dipartimento di Ingegneria Elettrica Elettronica e Informatica, University of Catania, 95029 Catania, Italy.}
\thanks {M. Frasca is with Dipartimento di Ingegneria Elettrica Elettronica e Informatica, University of Catania, 95029 Catania, Italy and with CNR-IASI, Institute for Systems Analysis and Computer Science ''A. Ruberti'', 00185 Rome, Italy (e-mail: mattia.frasca@dieei.unict.it).}
\thanks{F. Sorrentino, is with Department of Mechanical Engineering, Univ. of New Mexico, Albuquerque, New Mexico 87131, USA,
and with Dep. of Electrical and Computer Engineering, Univ. of New Mexico, Albuquerque, New Mexico 87131, USA}
\thanks{Louis M. Pecora is with U.S. Naval Research Laboratory, Washington, DC 20375, USA}
\thanks{S. Boccaletti is with CNR, Institute of Complex Systems, Via Madonna del Piano 10, 50019 Florence, Italy, with Unmanned Systems Research Institute, Northwestern Polytechnical University, Xi'an 710072, China and with Moscow Institute of Physics and Technology, Dolgoprudny, Moscow Region, 141701, Russian Federation.}
}

%


\maketitle

\begin{abstract}
Symmetries are an essential feature of complex networks as they regulate how the graph collective dynamics organizes into clustered states. We here show how to control network symmetries, and how to enforce patterned states of synchronization with nodes clustered in a desired way. Our approach consists of perturbing the original network connectivity, either by adding new edges or by adding/removing links together with modifying their weights. By solving suitable optimization problems, we furthermore guarantee that changes made on the existing topology are minimal. The conditions for the stability of the enforced pattern are derived for the general case, and the performance of the method is illustrated with paradigmatic examples. 
Our results are relevant to all the practical situations in which coordination of the networked systems into diverse groups may be desirable, such as for teams of robots, unmanned autonomous vehicles, power grids and central pattern generators.
\end{abstract}

\section{Introduction}\label{sec:introduction}

Control of complex networks is a challenging problem in many applications such as transport networks, power grids, wireless sensor networks, multi-robot teams, epidemics, social and biological systems \cite{mesbahi2010graph,liu2016control}. Steering the state variables associated with the network nodes in an efficient way requires techniques which often need to be tailored depending on the target of the control, the system size, or the characteristics of the network topology \cite{liu2011controllability,motter2004cascade,lombana2014distributed,lindmark2018role}. In particular, when the networked units feature an oscillatory dynamics, diverse problems arise depending on whether the control aims at achieving a collective behavior \cite{sorrentino2007controllability} or a clustered state \cite{gambuzza2019distributed}.
Synchronization, which is ubiquitous in natural and man-made networks \cite{boccaletti2002synchronization,pikovsky2003synchronization,boccaletti2018synchronization,aminzare2014synchronization,belykh2000hierarchy,gushchin2016phase}, can be in fact exhibited either as a global state in which all units follow the same trajectory or via patterned states where the system splits into subsets of units synchronized to each other. The latter phenomenon, known as cluster synchronization (CS), has been widely investigated both theoretically \cite{belykh2003persistent,lu2010cluster,russo2011symmetries,pecora2014cluster,sorrentino2016complete} and experimentally \cite{vardi2012synchronization,williams2013experimental,totz2015phase}. The key role in determining the composition of the clusters is played by the symmetries inherent to the network structure of interactions \cite{russo2011symmetries,pecora2014cluster,sorrentino2016complete}, or by equitable partitions of the network nodes \cite{schaub2016graph,gambuzza2019criterion}. With this in mind, algorithms for the generation of networks with symmetries \cite{klickstein2018generating,klickstein2018generating2}, and for optimization of synchronizability \cite{hart2019topological} have been proposed.

In this paper, we solve a practical and relevant problem: given an arbitrary network and a desired, arbitrary, set of symmetries, what is the minimal topological perturbation one can impose in order to control the given symmetries and obtain the associated CS state?
Depending on the way the perturbation strength is quantified (whether by the $L_2$-norm of the matrix representing the changes made, or by also taking into account the overall number of added/removed links) and the type of action performed, i.e., addition/removal or only addition of links, several solutions are provided via the formulation of different optimization problems. Our results may be used to control synchronization in arbitrary subsets of the network nodes, where one has simply to retrieve the proper symmetries associated to the desired subset and then apply our approach. In principle, there are two ways in which our approach can be used: 1) offline, i.e., by designing the network so as to enforce the formation of synchronous clusters; or 2) online, i.e., by introducing a cyber layer of controllers that provide virtual links corresponding to inputs equivalent to those associated to physical connections or canceling the effects of the existing ones.

Another motivating example for this work is the emergence of widespread synchronization in the human brain, which is associated to a pathological state such as epilepsy \cite{breakspear2005unifying}, whereas synchronization of little groups of neurons is essential for performing or optimizing specific cognitive and/or information-processing tasks \cite{varela2001brainweb}.

Earlier attempts towards topological control of CS have considered: i) the problem of stabilizing a synchronous pattern associated to a symmetry which already exists in the network \cite{lin2016controlling}; or ii) the control of specific small structures \cite{fu2013topological} or of a specific cluster state arising in networks of periodic oscillators \cite{lehnert2014controlling}; iii) the control through symmetries of a single subgroup of nodes \cite{gambuzza2019distributed,ursino2019control}. All these studies do not provide a systematic method for inducing arbitrary desired symmetries in a network, which is instead the main focus of our paper.

\section{Preliminaries and problem formulation}

A graph $\mathcal{G}$ is defined by the set of vertices (or nodes) $\mathcal{V}(\mathcal{G}) = \{1, . . . , N\}$ and the set of edges (or links) $\mathcal{E}(\mathcal{G}) \subseteq \mathcal{V} \times \mathcal{V}$. An edge from node $i$ to node $j$ is an ordered pair $(i, j)$. The graph is said to be undirected if for any $(i, j)\in \mathcal{V}$, then $(j,i)\in \mathcal{V}$. In the rest of the paper, we will focus on undirected graphs, without self-loops (i.e., links starting and ending in the same node).

To represent the graph $\mathcal{G}$, its adjacency and Laplacian matrices will be used. The adjacency matrix $\mathrm{A}$ is an $N\times N$ matrix of constant coefficients, where $\mathrm{A}_{ii}=0$ (as there are no self-loops), $\mathrm{A}_{ij}=0$ if units $i$  and $j$ are not connected, and $\mathrm{A}_{ij}=\mathrm{A}_{ji}>0$ if there is a link between $i$ and $j$ (with the value of $\mathrm{A}_{ij}$ encoding the link weight). The Laplacian matrix is a $N\times N$ positive semidefinite, zero row-sum matrix with entries $\mathrm{L}_{ij}=-\mathrm{A}_{ij}$ for $i \neq j$ and $\mathrm{L}_{ii}=\sum\limits_{j=1}^N \mathrm{A}_{ij}$. For connected graphs, the eigenvalues of $\mathrm{L}$ can be ordered as: $0=\lambda_1 < \lambda_2 \leq \lambda_3 \leq \  \ldots  \leq \ \lambda_N$.

To each node, we associate a dynamical system, which interacts with the other units through the edges of the network. We will refer to such a system as a network of coupled dynamical units. In particular, we consider identical dynamical units, described by the following equations:
\begin{equation}
\label{eq:eq1Lapl}
\dot{\mathbf{x}}_i=\mathbf{f}(\mathbf{x}_i)-\sigma\sum_{j=1}^N\mathrm{L}_{ij} \mathbf{h}(\mathbf{x}_j) ,
\end{equation}
\noindent where $i=1,\ldots,N$,  $\mathbf{x}_i \in \mathbb{R}^n$ is the state vector of the $i$-th unit, $\mathbf{f}:\mathbb{R}^n\rightarrow \mathbb{R}^n$ ($\mathbf{h} :\mathbb{R}^n\rightarrow \mathbb{R}^n$) is the function describing the evolution of the uncoupled systems (the coupling function), and  $\sigma$ is the coupling strength.

Two nodes of the network are said to be synchronized if $\lim\limits_{t\rightarrow \infty} \| \mathbf{x}_j-\mathbf{x}_i\|=0$; the entire network is said to be globally synchronized if this condition holds for any pair of nodes. Whether or not Eqs.~(\ref{eq:eq1Lapl}) exhibit global synchronization (GS) depends on the unit dynamics, the network topology, and the coupling strength \cite{barahona2002synchronization,boccaletti2008synchronized}. {Precisely, application of the Master Stability Function approach \cite{pecora1998master,boccaletti2008synchronized} to stability of the GS state yields a classification of the systems into three classes. For class I systems, the maximum Lyapunov exponent of the modes transverse to the GS manifold, indicated as $\Lambda_{max}$, is always positive and GS is always unstable for any coupling. For class II systems, there is a threshold value $\alpha_1$ such that $\Lambda_{max}(\alpha)<0$ for $\alpha > \alpha_1$; under these circumstances Eqs.~(\ref{eq:eq1Lapl}) admit a stable GS manifold (${\mathbf{x}}_i(t)={\mathbf{x}}_j(t)$ $\forall i,j$) if $\sigma > \alpha_1/\lambda_2$. For class III systems, $\Lambda_{max}(\alpha)<0$ holds in a finite interval of values, i.e., in $\alpha \in [\alpha_1,\alpha_2]$, and Eqs.~(\ref{eq:eq1Lapl}) admit a stable GS manifold if $\alpha_1/\lambda_2 < \sigma < \alpha_2/\lambda_N$}. $\alpha_1$ and $\alpha_2$ depend on $\mathbf{f}$ and $\mathbf{h}$, and are suitable intersections between the abscissa axis and the function representing the maximum of all Lyapunov exponents transverse to the synchronous manifold vs. the variable $\alpha=\sigma \lambda$ \cite{barahona2002synchronization,boccaletti2008synchronized}.

In networks with symmetries, patterns of synchronized units, i.e., cluster synchronization (CS), may also emerge, with the nodes clustering in groups with synchronous behavior, generally distinct from that of the other groups. Indicating with $M$ the number of groups, CS occurs when the nodes cluster into the sets $V_1,\ldots,V_M$ with $V_h\cap V_l=\emptyset$ and $\bigcup\limits_{h=1}^MV_h=\mathcal{V}$ such that $\lim\limits_{t\rightarrow\infty}\| \mathbf{x}_j-\mathbf{x}_i\|=0$, for any $i,j\in V_h$, $h=1,\ldots,M$. Formally, an object is said to have a symmetry if there exists an operation that, when applied to it, leaves it unchanged \cite{heine2007group}. For graphs, symmetries are associated to automorphisms, i.e. permutations of the nodes that preserve the connectivity pattern \cite{latora2017}. All symmetries of a graph form a mathematical group, where each element is represented as a square permutation matrix $\mathrm{R}^g = (r_{ij})$, with $r_{ij}=1$ if node $j$ is mapped to node $i$ under the permutation, and $r_{ij}=0$ otherwise.
The symmetry group induces a partition of the nodes into disjoint sets called orbits. Orbits include all nodes that get mapped into each other after application of all  symmetries of the group. Nodes that are in the same orbit may cluster-synchronize because the equations of motion are equivariant. This forms natural CS patterns, where the exact composition and stability of clusters can be determined with group-theoretical considerations \cite{pecora2014cluster}.

In this paper, we investigate the problem of changing, in a minimal way, a network so as to impose a given set of symmetries and stabilizing the associated synchronization pattern. The problem considered is hence twofold. 1) Given a network, described by its adjacency matrix $\mathrm{A}$, and a target set of symmetries, the first problem is to find a perturbation $\mathrm{\Delta A}$ such that the new network, with adjacency matrix $\mathrm{A}+\mathrm{\Delta A}$ admits the given set of symmetries. 2) Given a network of identical dynamical units as in (\ref{eq:eq1Lapl}), after introducing in the connectivity the perturbation $\mathrm{\Delta A}$ found solving the first problem, one obtains a synchronization pattern associated to the symmetries of the new structure; the second problem is to stabilize the desired CS pattern through a proper setting of the coupling strength $\sigma$.

Notice that, although developed in the context of CS, the solution to the first problem may be of more general interest and be applicable to all scenarios where inducing symmetries in a network is desirable.

\section{Main results}

This section illustrates the solution proposed for the problem of inducing network symmetries and a stable cluster state in network (\ref{eq:eq1Lapl}), and is articulated in three parts. First, we show that the problem of inducing symmetries in a network of $N$ nodes via perturbations of the original structure can be recast in terms of a Sylvester equation, which, in turn, may be rewritten as a linear system in a set of $N(N-1)/2$ unknowns. Second, we formulate three different optimization problems to compute the perturbations and, hence, to induce the target network symmetries. Third, we study the stability of the target CS state to determine the suitable range for coupling.

\subsection{A Sylvester equation for inducing network symmetries}
\label{sec:sylvester}

We discuss here the method we propose to modify in a minimal way a generic network, in order to produce a new structure that has a given set of symmetries. To do so, we either introduce new links in the topology, or remove some of the existing ones, or change the link weights. Let us indicate with $\mathrm{\bar{A}}$ the adjacency matrix of the network generated after application of the perturbation $\mathrm{\Delta A}$ and write it as $\mathrm{\bar{A}}= \mathrm{A} + \mathrm{\Delta A}$, where $\mathrm{A}$ is the adjacency matrix of the pristine network. Furthermore, let $\{\mathrm{R}_1,\mathrm{R}_2, \ldots, \mathrm{R}_Q\}$ be a set of permutations in the desired symmetry group $\mathcal{H}$.

In order to design a small perturbation $\mathrm{\Delta A}$, we use the property that $\mathrm{\bar{A}}$ and each of the permutation matrices $\mathrm{R}_i$ ($i=1,\ldots,Q$) have to commute \cite{heine2007group}, i.e.,

\begin{equation}
\label{eq:LyapEq0}
\mathrm{R_i}\mathrm{\bar{A}}=\mathrm{\bar{A}}\mathrm{R_i} \ .
\end{equation}

Substituting the expression for $\mathrm{\bar{A}}$, one has:
\begin{equation}
\label{eq:LyapEqV2}
\mathrm{R_i}\cdot\mathrm{\Delta A}-\mathrm{\Delta A}\cdot\mathrm{R_i}=-\mathrm{R_i}\mathrm{A}+\mathrm{A}\mathrm{R_i} \ .
\end{equation}

Notice that $\mathrm{R}_1,\mathrm{R}_2, \ldots, \mathrm{R}_Q$ do not have to be all the symmetries of the group, but it is sufficient to only consider the generators of the group. To show this, let $\mathrm{R}_p$ be a generic symmetry of the group which can be expressed as the composition of the product of two (or more) generators.
Without lack of generality, let us suppose that $\mathrm{R}_p=\mathrm{R}_i\mathrm{R}_j$ for given $i$ and $j$.
Then, one has to show that if $\mathrm{\Delta A}$ is such that $\mathrm{\bar{A}}=\mathrm{A}+\mathrm{\Delta A}$ satisfies

\begin{equation}
\label{eq:conRi}
\mathrm{R}_i\mathrm{\bar{A}}-\mathrm{\bar{A}}\mathrm{R}_i=0
\end{equation}

\noindent and

\begin{equation}
\label{eq:conRj}
\mathrm{R}_j\mathrm{\bar{A}}-\mathrm{\bar{A}}\mathrm{R}_j=0,
\end{equation}

\noindent then

\begin{equation}
\label{eq:conRp}
\mathrm{R}_p\mathrm{\bar{A}}-\mathrm{\bar{A}}\mathrm{R}_p=0
\end{equation}

\noindent is also true.

This follows from

\begin{equation}
\label{eq:LyapEqComp}
\mathrm{R}_p\mathrm{\bar{A}}-\mathrm{\bar{A}}\mathrm{R}_p=\mathrm{R}_i\mathrm{R}_j\mathrm{\bar{A}}-\mathrm{\bar{A}}\mathrm{R}_i\mathrm{R}_j.
\end{equation}

Thanks to Eq. (\ref{eq:conRi}), this can be rewritten as

\begin{equation}
\label{eq:LyapEqComp2}
\mathrm{R}_i\mathrm{R}_j\mathrm{\bar{A}}-\mathrm{\bar{A}}\mathrm{R}_i\mathrm{R}_j=\mathrm{R}_i(\mathrm{R}_j\mathrm{\bar{A}}-\mathrm{\bar{A}}\mathrm{R}_j),
\end{equation}

\noindent from which, in virtue of Eq. (\ref{eq:conRj}), Eq. (\ref{eq:conRp}) follows. The same argument applies to permutations obtained as the product of more than two generators.

Equation~(\ref{eq:LyapEqV2}) is a Sylvester equation\footnote{Note that, for $\mathrm{R_i}=\mathrm{I}$, Eq.~(\ref{eq:LyapEqV2}) becomes a trivial identity, as the identity matrix represents a suitable symmetry for any network.} that can be conveniently recast by vectorization \cite{macedo2013typing} to obtain $\mathfrak{R}_i\cdot \mathrm{vec}(\Delta {A})= \mathrm{vec}(-\mathrm{R_i}\mathrm{A}+\mathrm{A}\mathrm{R_i})$, i.e. by that linear transformation which converts a matrix $\mathrm{C}\in \mathbb{R}^{m\times n}$ into a column vector $\mathrm{vec}(\mathrm{C})$, corresponding to parsing $\mathrm{C}$ in column-major order, i.e.,
$\mathrm{vec}(\mathrm{C})=\left[C_{11}~C_{21}~ \ldots ~ C_{m1} ~ \ldots ~ C_{1n} ~ C_{2n} ~ \ldots ~ C_{mn}\right]^T$.
By doing so, one obtains:
\begin{equation}
\label{eq:LyapEqVec}
\mathfrak{R}_i\cdot \mathrm{vec}(\Delta {A})= \mathrm{vec}(-\mathrm{R_i}\mathrm{A}+\mathrm{A}\mathrm{R_i}) ,
\end{equation}
\noindent
where $\mathfrak{R}_i=\mathrm{I}_N\otimes\mathrm{R_i}-\mathrm{R_i}^T\otimes\mathrm{I}_N$
with $i=1,\ldots,Q$. To simultaneously satisfy these conditions, let us define $\mathfrak{R}=\left [  \mathfrak{R}_1, \ \mathfrak{R}_2, \ \ldots , \ \mathfrak{R}_Q  \right ]^T$
and $\mathrm{B}=[  \mathrm{vec}(-\mathrm{R_1}\mathrm{A}+\mathrm{A}\mathrm{R_1}), \ \mathrm{vec}(-\mathrm{R_2}\mathrm{A}+\mathrm{A}\mathrm{R_2}), \ \ldots , \ \mathrm{vec}(-\mathrm{R_Q}\mathrm{A}+\mathrm{A}\mathrm{R_Q})  ]^T$, the problem is then reformulated in terms of finding the solution of an algebraic linear system

\begin{equation}
\label{eq:main}
\mathfrak{R}\cdot \mathrm{vec}(\mathrm{\Delta A})=\mathrm{B}.
\end{equation}

Since the complete graph $\mathcal{K}_N$ is symmetric with respect to all possible permutations, adding links to the original topology until complete connectivity is reached always provides a trivial solution.
Such a solution is however highly inefficient, as it requires adding the largest possible number of new links. In the next section, we present several methods, formulated as distinct optimization problems, to solve Eq. (\ref{eq:main}) in a more efficient way.

\subsection{Optimization-based solutions}
\label{sec:optmethods}

We propose three methods to solve Eq. (\ref{eq:main}), differing for the additional contraints that can be incorporated in the optimization problem and that yield networks with different characteristics. They are illustrated in the following.

\emph{Moore-Penrose Inverse (MPI).} The first method, referred to as \emph{Moore-Penrose Inverse (MPI)}, is based on solving the optimization problem:

\begin{equation}
\min \| \mathrm{vec}(\mathrm{\Delta A}) \|_2, \mathrm{~subject~to~}\mathfrak{R}\cdot \mathrm{vec}(\mathrm{\Delta A})=\mathrm{B}
\end{equation}

Here, by minimizing the $L_2$ norm of $\mathrm{vec}(\mathrm{\Delta A})$, one seeks a solution with minimum changes performed on the link weights. The problem admits an analytical solution given by:

\begin{equation}
\label{eq:sol1}
\mathrm{vec}(\mathrm{\Delta A})=\mathfrak{R}^\dag\mathrm{B}.
\end{equation}

\noindent where $\mathfrak{R}^\dag$ is the Moore-Penrose inverse of $\mathfrak{R}$. The result is generically a weighted graph.

\emph{Lasso (least absolute shrinkage and selection operator).} The second method considers the following optimization problem:

\begin{equation}
\min \{ \| \mathfrak{R}\cdot \mathrm{vec}(\mathrm{\Delta A})-\mathrm{B} \|^2_2+\beta\|  \mathrm{vec}(\mathrm{\Delta A}) \|_1 \}
\end{equation}

\noindent where $\beta>0$ is a regularization parameter weighting the two terms to optimize. The objective function now includes a first term $\| \mathfrak{R}\cdot \mathrm{vec}(\mathrm{\Delta A})-\mathrm{B} \|^2_2$ whose minimization warrants that the new network admits the desired set of symmetries and a second term $\|  \mathrm{vec}(\mathrm{\Delta A}) \|_1$ whose minimization warrants the sparsity of the solution, i.e., that a minimum number of links is added/removed with respect to the original structure. The problem is solved through lasso, a convex optimization method based on compressive sensing techniques \cite{han2015robust} and available in many softwares for mathematical computation. For this reason, this approach is referred to as the \emph{lasso} method.

\emph{Connectedness preserving optimization (CPO).} In the third method, we consider that the only allowed change to the original structure is to add links. To this aim, we impose $\mathrm{\Delta A}_{ij}\geq 0$ for $i\neq j$ and rewrite the optimization problem as: 

\begin{equation}
\label{eq:thirdmethod}
\begin{array}{l}
\min \| \mathrm{vec}(\mathrm{\Delta A}) \|_2,\\
\mathrm{~subject~to~}\mathfrak{R}\cdot \mathrm{vec}(\mathrm{\Delta A})=\mathrm{B} \mathrm{~and~to~}\mathrm{\Delta A}_{ij}\geq 0, \forall i\neq j
\end{array}
\end{equation}

This problem is solved through linear programming. Because it preserves the property of the network to be connected if the original structure is connected, we refer to it as as \emph{connectedness preserving optimization (CPO)}. 

For small-size networks, the problem can be recast with integer variables and thus solved with integer linear programming such that, if the pristine network is unweighted, so is the controlled one. To do this, one considers that the elements of $\mathrm{\Delta A}$ can assume only two values, $\mathrm{\Delta A}_{ij}=1$ if a new link $(i,j)$ is added, $\mathrm{\Delta A}_{ij}=0$ otherwise. The optimization problem is formulated as:

\begin{equation}
\label{eq:CPOinteger}
\min \mathbf{c}^T \mathbf{q}, \mathrm{~subject~to~}\mathfrak{R} \mathbf{q}=\mathrm{B}
\end{equation}

\noindent where $q_h$ ($h=1,\ldots,N^2$) are binary $\{0,1\}$ variables and $\mathbf{c}$ is a vector with $c_h=0$ if $h=\{1,N+1,2N+2,\ldots,N^2\}$ and $c_h=1$, otherwise. $q_h=1$ means that $\mathrm{\Delta A}_{ij}=1$, i.e., a link $(i,j)$ with $i=\mod(h,N)$ and $j=[h/N]$ is added to the original network; $q_h=0$ implies that $\mathrm{\Delta A}_{ij}=0$. Minimization of the objective function $\mathbf{c}^T \mathbf{q}$ corresponds to adding the minimum number of links to the original structure, as each new link contributes a unity to the term $\mathbf{c}^T \mathbf{q}$ (self-loops are excluded as $c_h=0$ if $h=\{1,N+1,2N+2,\ldots,N^2\}$).

As the standard routines available for integer linear programming do not scale well with the size of the problem, for the analysis of large networks the formulation (\ref{eq:thirdmethod}) is preferred.

\begin{figure*}
\begin{center}
\subfigure[]{\includegraphics[width=0.23\textwidth]{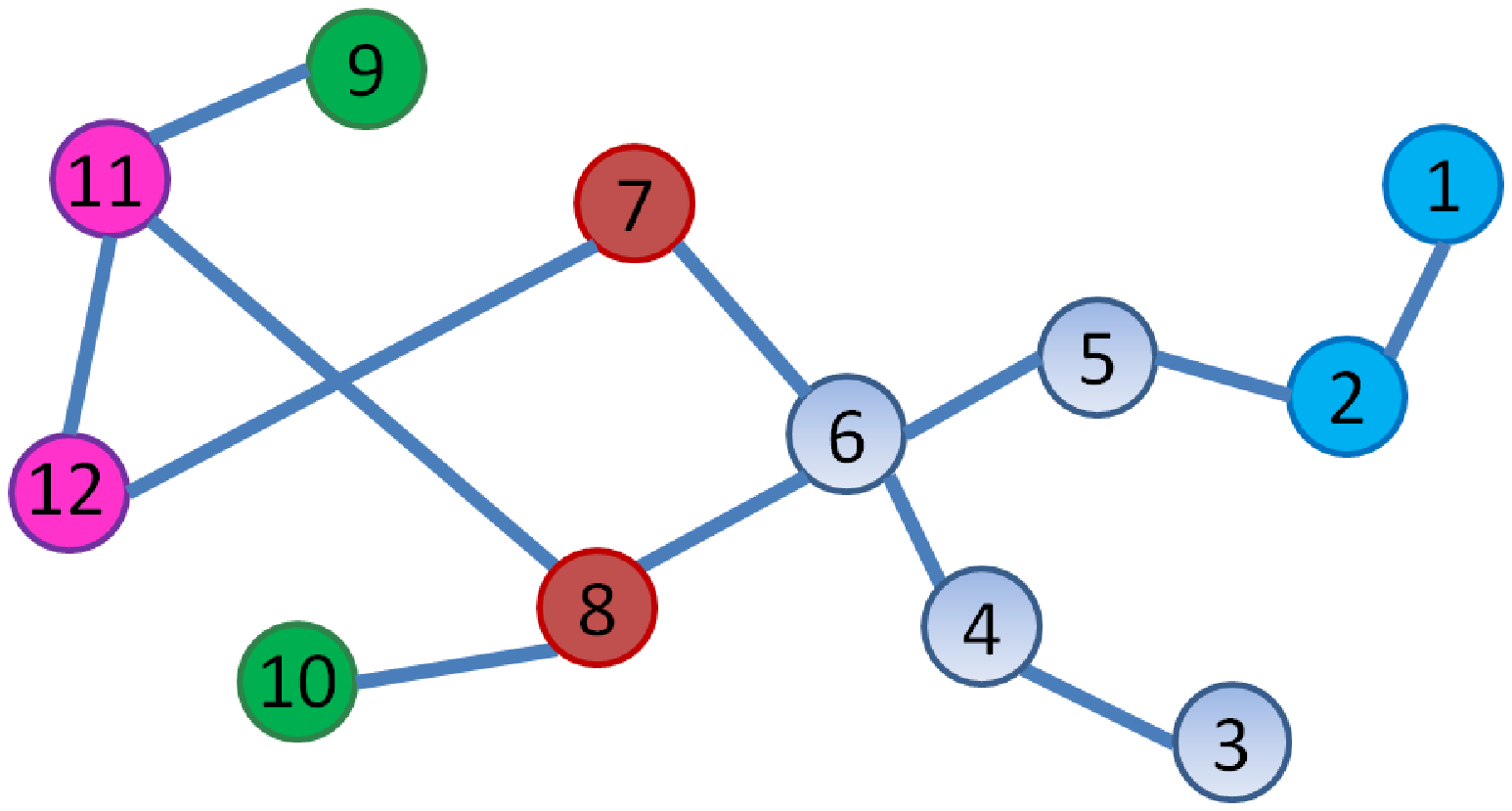}\label{fig:Ex1A}}
\subfigure[]{\includegraphics[width=0.23\textwidth]{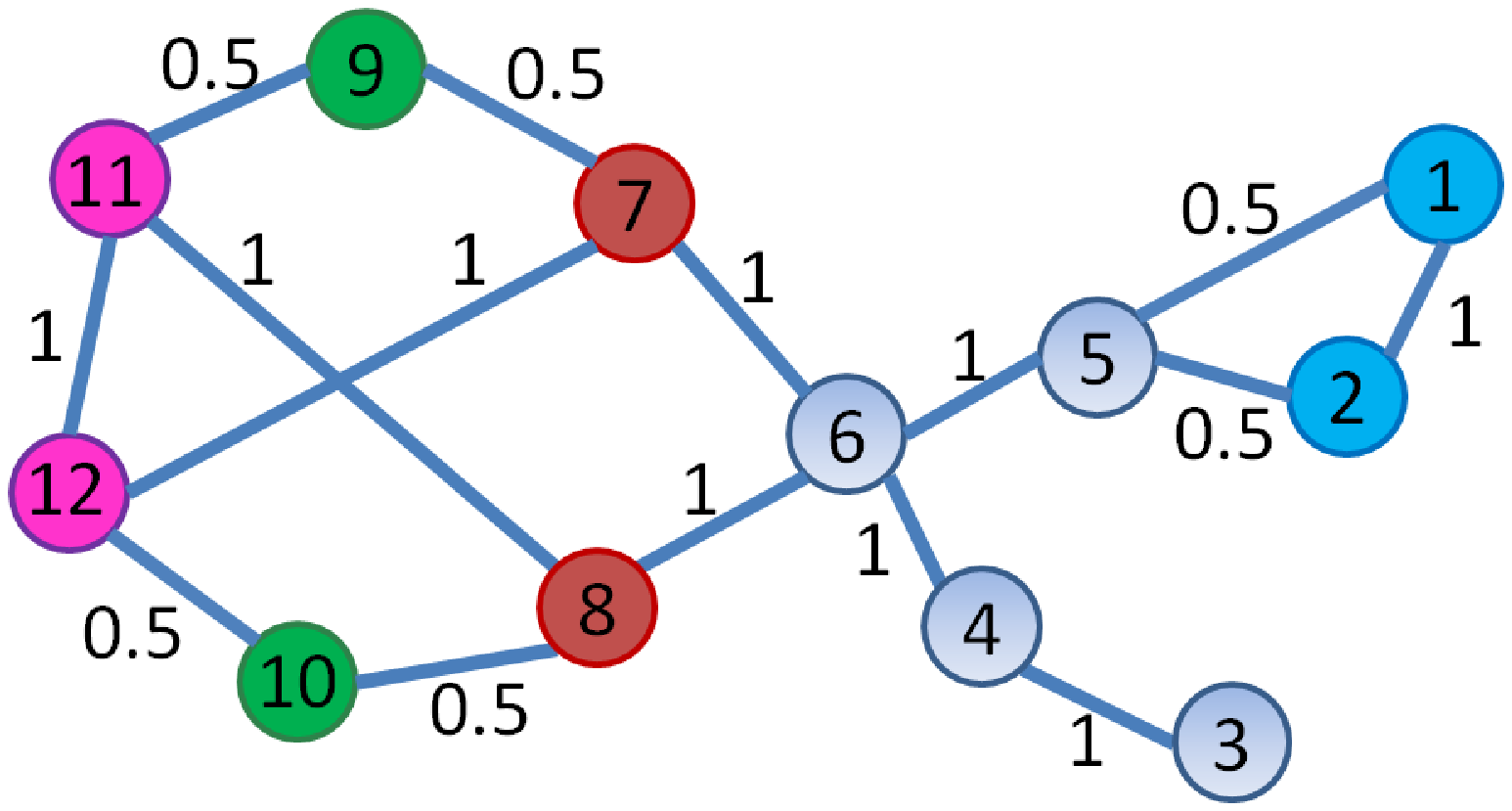}\label{fig:Ex1B}}
\subfigure[]{\includegraphics[width=0.23\textwidth]{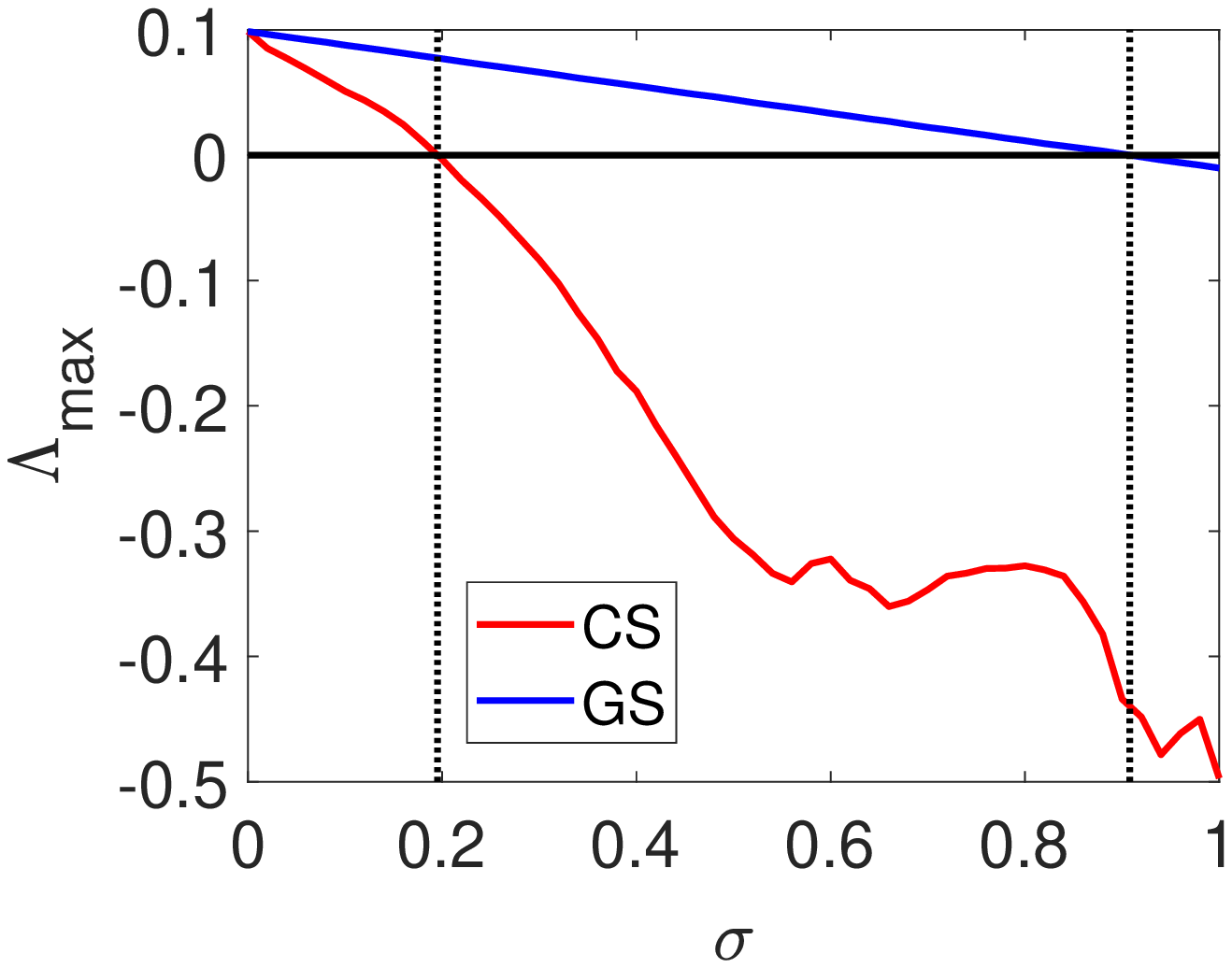}\label{fig:Ex1C}}
\subfigure[]{\includegraphics[width=0.23\textwidth]{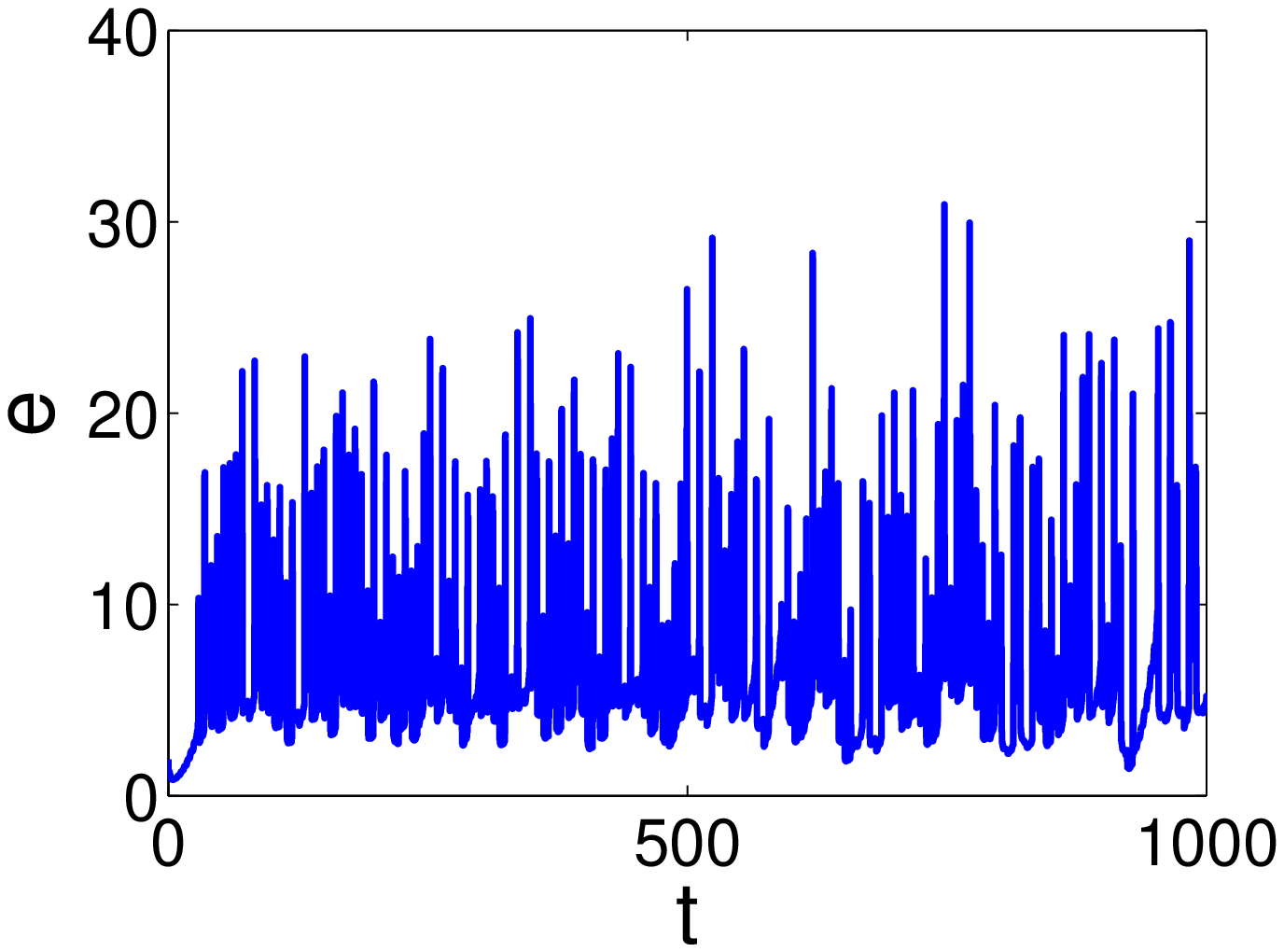}\label{fig:Ex1D}}
\subfigure[]{\includegraphics[width=0.23\textwidth]{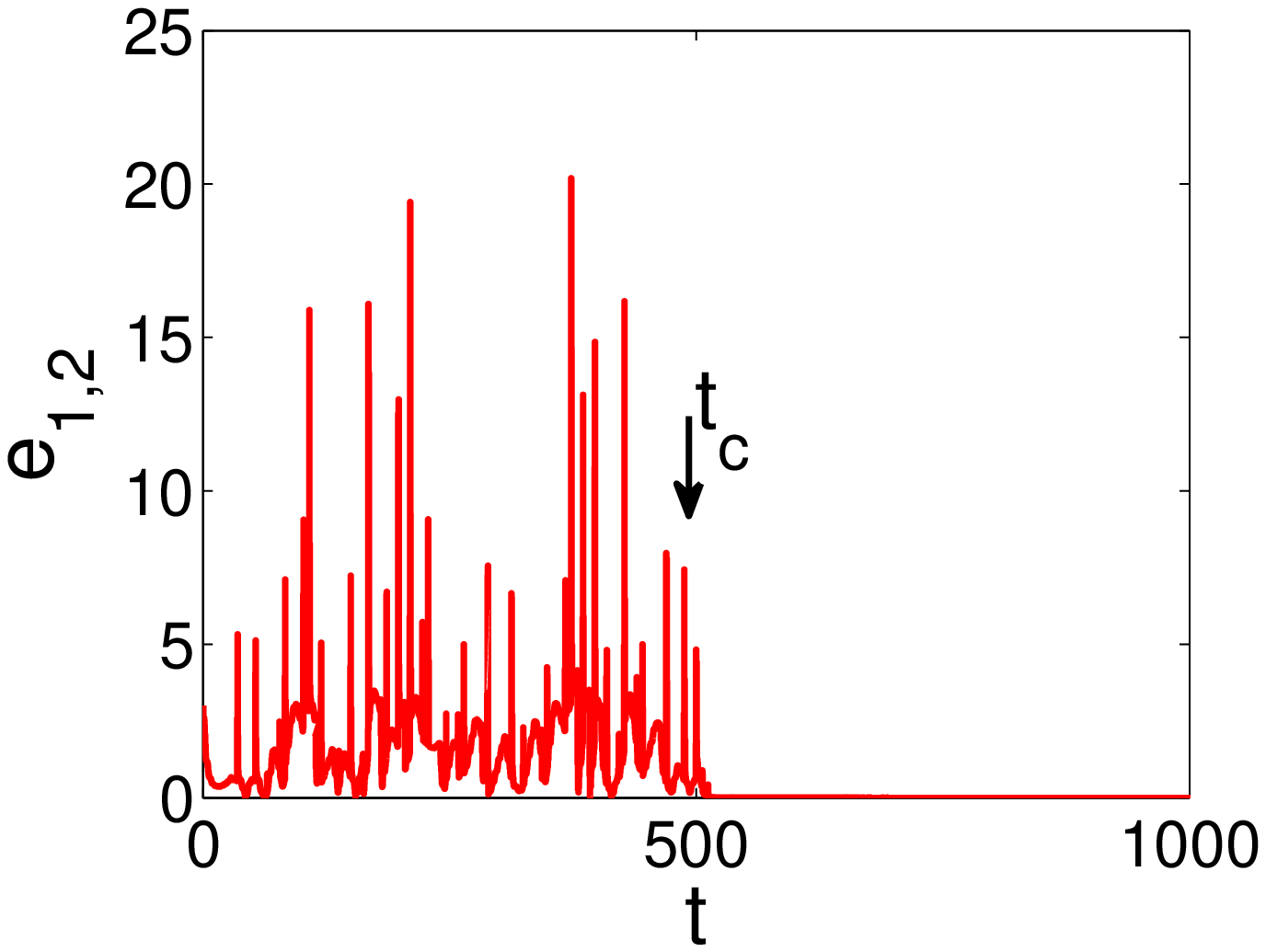}\label{fig:Ex1E}}
\subfigure[]{\includegraphics[width=0.23\textwidth]{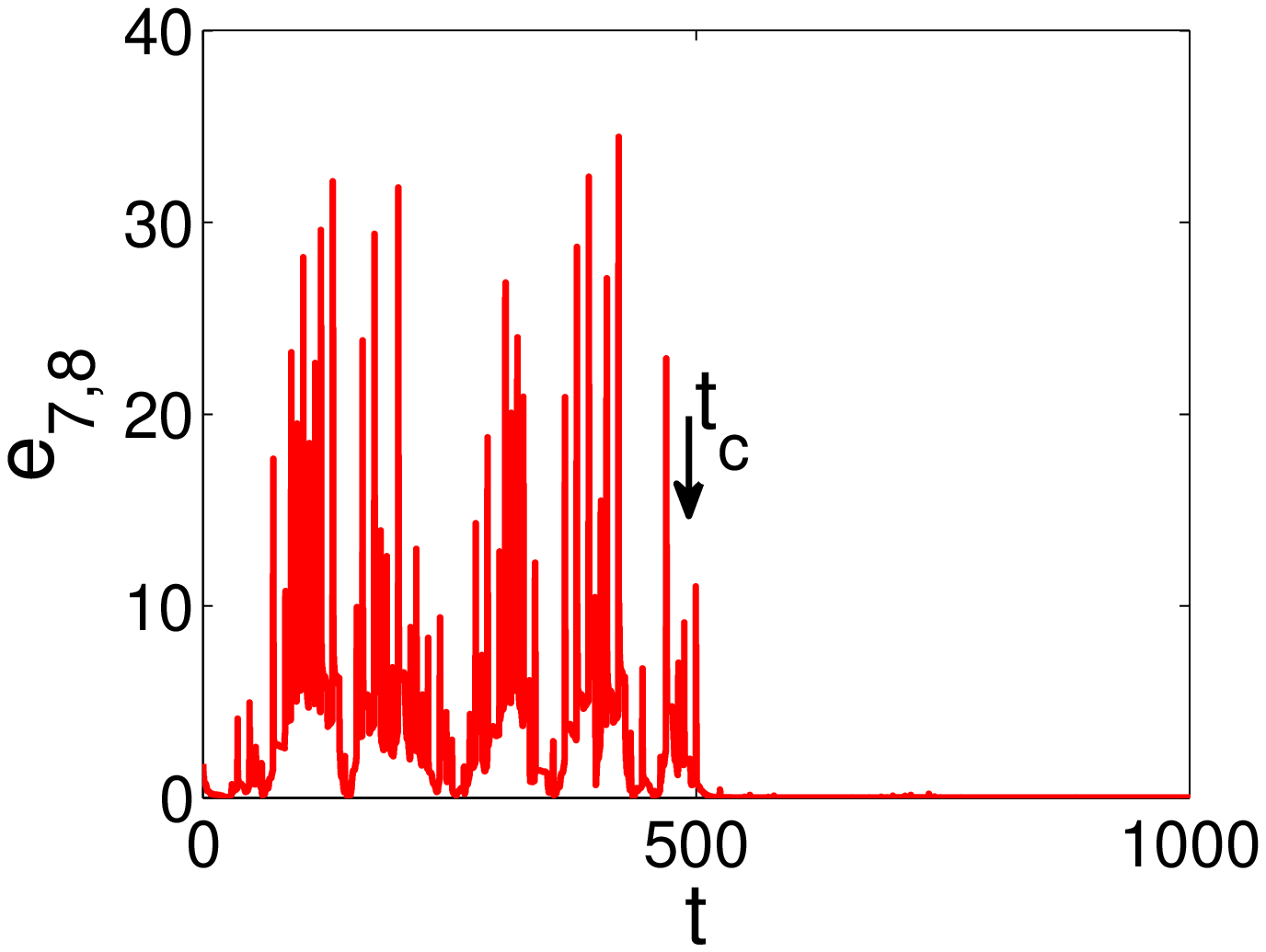}\label{fig:Ex1F}}
\subfigure[]{\includegraphics[width=0.23\textwidth]{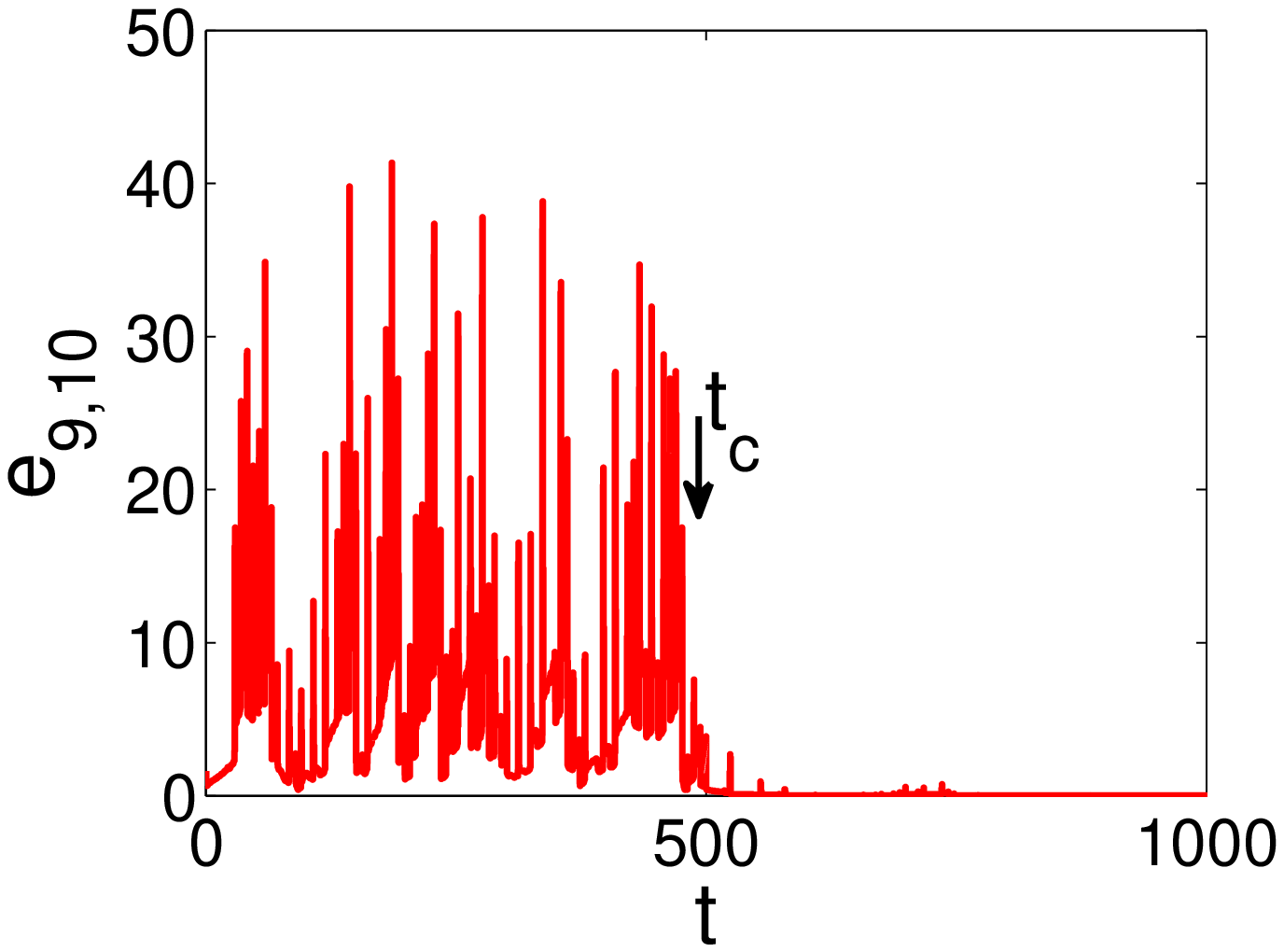}\label{fig:Ex1G}}
\subfigure[]{\includegraphics[width=0.23\textwidth]{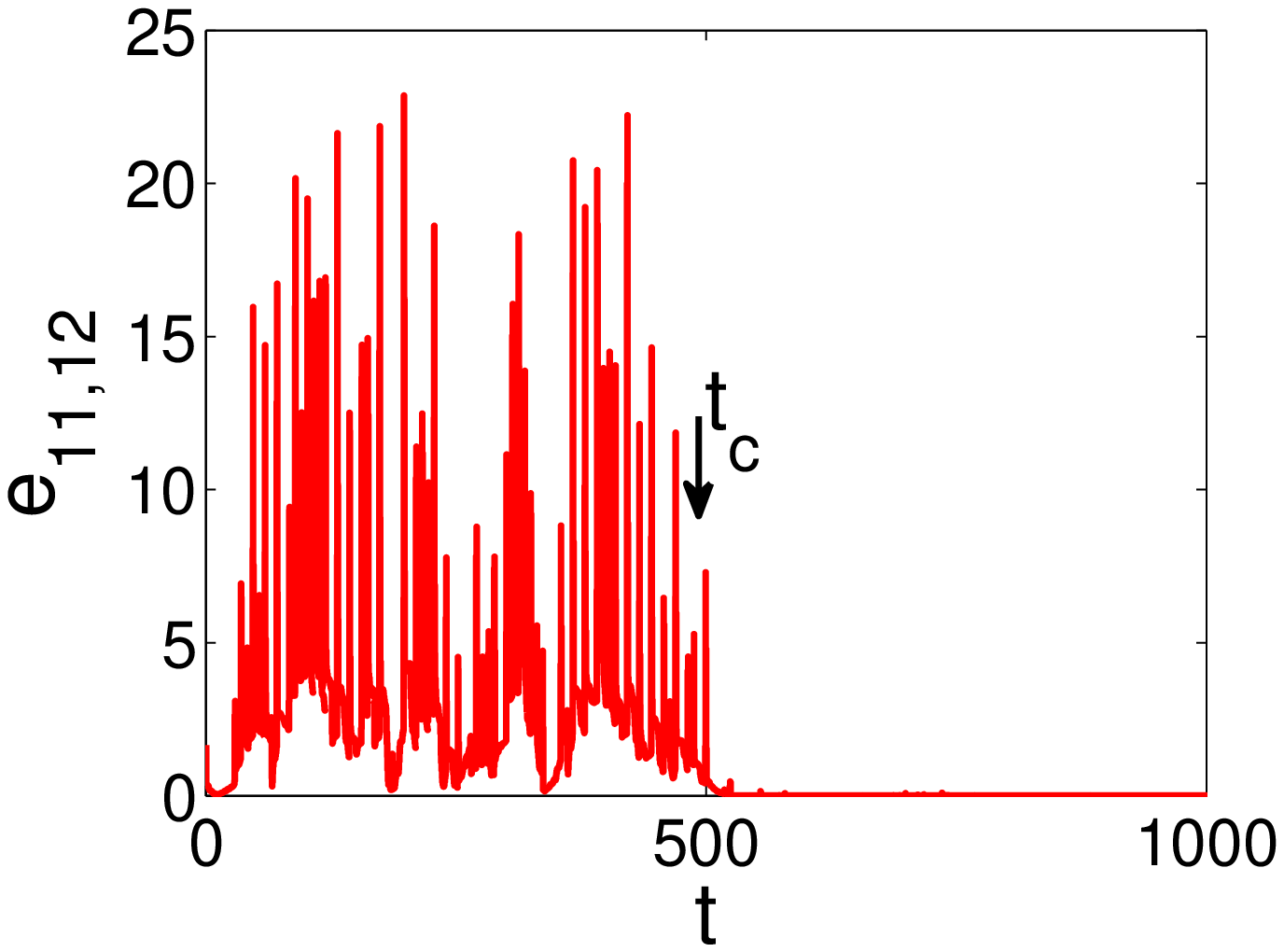}\label{fig:Ex1H}}
\caption{\label{fig:Ex1} {Controlling cluster synchronization}. (a) Sketch of the pristine network topology, with no symmetries; (b) controlled weighted network, with units 1 and 2 (blue nodes), 7 and 8 (brown nodes), 9 and 10 (green nodes), and 11 and 12 (pink nodes) forming the distinct clusters $V_1,V_2,V_3,V_4$; (c) maximum transverse Lyapunov exponent, $\Lambda_{max}(\sigma)$, for CS and GS. Panels (d-h) report the time evolution of the synchronization error $e$ for GS (d), and of the errors $e_{1,2}$ (e), $e_{7,8}$ (f), $e_{9,10}$ (g), and $e_{11,12}$ (h). Results in panels (d-h) refer to $\sigma=0.25$. The vertical arrows in panels (e-h) labeled with $t_c$ indicate the instant of time at which the pristine network is substituted with the controlled one in the simulations.}
\end{center}
\end{figure*}

\subsection{Stability of cluster synchronization}

Let $\bar{\mathrm{L}}$ be the Laplacian matrix of the new network, i.e., $\bar{\mathrm{L}}=\bar{\mathrm{D}}-\bar{\mathrm{A}}$, where $\bar{\mathrm{D}}$ is the diagonal matrix containing the node degrees (or the node strengths) of the new network. Replacing $\bar{\mathrm{L}}$ in Eqs. (\ref{eq:eq1Lapl}) and rewriting in compact form, one has
\begin{equation}
\label{eq:compact}
\dot{\mathbf{x}}=\mathbf{F}(\mathbf{x})-\sigma(\bar{\mathrm{L}}\otimes\mathrm{I}_n)\mathbf{H}(\mathbf{x}),
\end{equation}
\noindent with $\mathbf{x}=[\mathbf{x}_1^T,\mathbf{x}_2^T,\ldots,\mathbf{x}_N^T]^T$,
$\mathbf{F}(\mathbf{x}) = [\mathbf{f}^T(\mathbf{x}_1), \mathbf{f}^T(\mathbf{x}_2), \ldots, \mathbf{f}^T(\mathbf{x}_N)]^T$, and
$\mathbf{H}(\mathbf{x}) = [\mathbf{h}^T(\mathbf{x}_1), \mathbf{h}^T(\mathbf{x}_2), \ldots, \mathbf{h}^T(\mathbf{x}_N)]^T$.
Notice that the resulting network can also admit other (i.e., not imposed) orbits or equitable partitions, and, in case of class II systems, will certainly display GS at sufficiently large coupling strength. The controlled symmetries will ultimately determine the division into clusters of the network nodes. Let $M$ be the number of distinct clusters, giving rise to $M$ trajectories $\mathbf{s_1}(t), \mathbf{s_2}(t), \ldots, \mathbf{s_M}(t)$, and let $V_l$ be the set of nodes belonging to cluster $l$, then the CS state will be characterized by $\mathbf{x_i}(t) =\mathbf{s}_l(t)$ $\forall i \in V_l$.

We now study the local stability of the CS state, rewriting the state variables as $\mathbf{x}_i=\mathbf{s}_l+\delta\mathbf{x}_i$ $\forall i\in V_l$, with $\delta\mathbf{x}_i$ being a small perturbation of node $i$ around $\mathbf{s}_l$. Linearizing Eqs. (\ref{eq:compact}), one obtains the dynamics of the perturbations:

\begin{equation}
\label{eq:networklinearized}
\mathbf{\delta \dot{x}}=\left [ \sum\limits_{l=1}^M\mathrm{E}^l\otimes \mathbf{Df}(\mathbf{s}_l)-\sigma\bar{\mathrm{L}}\sum\limits_{l=1}^M\mathrm{E}^l\otimes\mathbf{Dh}(\mathbf{s}_l) \right] \mathbf{\delta x},
\end{equation}

\noindent where $\mathbf{Df}$ ($\mathbf{Dh}$) is the Jacobian of $\mathbf{f}$ ($\mathbf{h}$) evaluated, for each cluster, around the synchronous solution $\mathbf{s}_l$. $\mathrm{E}^l$ is a diagonal matrix encoding the nodes belonging to cluster $l$, with $l=1, \ldots, M$, i.e., $\mathrm{E}^l_{ii}=1$ if node $i \in V_l$ and $\mathrm{E}^l_{ii}=0$ otherwise. $\mathbf{\delta x}=[\mathbf{\delta x}_1^T,\mathbf{\delta x}_2^T,\ldots,\mathbf{\delta x}_N^T]^T$ is the stack vector of the perturbations associated to the nodes.

The variational equation can be rewritten in a coordinate system where the Laplacian is block diagonal. This requires the calculation of a matrix $\mathrm{T}$ obtained from the computation of the irreducible representations of the symmetry group, usually performed through dedicated discrete algebra software. The matrix $\mathrm{T}$ allows the definition of new transformed variables as $\mathbf{\delta y}= \mathrm{T}^{-1} \otimes \mathrm{I}_n \mathbf{\delta x}$ such that Eqs. (\ref{eq:networklinearized}) become:
\begin{equation}
\label{eq:networklinearized2p}
{\mathbf{\delta \dot{y}}}=\left [ \sum\limits_{l=1}^M\tilde{\mathrm{E}}^l\otimes \mathbf{Df}(\mathbf{s}_l)-\sigma\tilde{\mathrm{L}}\sum\limits_{l=1}^M\tilde{\mathrm{E}}^l\otimes\mathbf{Dh}(\mathbf{s}_l) \right] \mathbf{\delta y},
\end{equation}

\noindent where $\tilde{\mathrm{E}}^l = \mathrm{T}^T \mathrm{E}^l \mathrm{T}$ and $\tilde{\mathrm{L}} =\mathrm{T} \bar{\mathrm{L}} \mathrm{T}^{-1}$, now block-diagonal.

More in details, $\tilde{\mathrm{L}}$ is partitioned in two main diagonal blocks:

\begin{equation}
\tilde{\mathrm{L}}= \left(\begin{array}{cc} \tilde{\mathrm{L}}_{||} & \mathrm{0}\\ \mathrm{0} & \tilde{\mathrm{L}}_{\bot} \end{array}\right).
\end{equation}

The structure of $\tilde{\mathrm{L}}$ and $\tilde{\mathrm{E}}^l$ allows one to obtain $M\times n$ equations associated to the block $\tilde{\mathrm{L}}_{||}$ (characterizing the motion parallel to the synchronous manifold), and $(N-M)\times n$ equations associated to  $\tilde{\mathrm{L}}_{\bot}$ (characterizing the motion transverse to the synchronous manifold).

Stability of the system requires that the transverse modes damp out, and to this aim the lower dimensional ODEs derived from the lower block of (\ref{eq:networklinearized2p}) are studied:

\begin{equation}
\label{eq:networklinearized2}
\mathbf{\delta \dot{y}}_{\bot}=\left [ \sum\limits_{l=1}^M\tilde{\mathrm{E}}^l\otimes \mathbf{Df}(\mathbf{s}_l)-\sigma \tilde{\mathrm{L}}_{\bot} \sum\limits_{l=1}^M\tilde{\mathrm{E}}^l\otimes\mathbf{Dh}(\mathbf{s}_l) \right] \mathbf{\delta y}_{\bot} \ ,
\end{equation}

\noindent where $\mathbf{\delta y}_{\bot}=[\mathbf{\delta y}^T_{N-M+1},\ldots,\mathbf{\delta y}^T_{N}]^T$. To assess the stability of the cluster synchronous state, the maximum Lyapunov exponent corresponding to the transverse blocks is computed from Eqs.~(\ref{eq:networklinearized2}) and stability is achieved if i) this quantity is negative, and ii) the synchronous pattern is asymptotically valid, i.e. it is observed after integrating the equations for a sufficiently long time. The latter condition stems from the fact that coarser cluster states (states where two or more clusters merge together) may exist in the same network structure, and the target cluster state may converge to one of them \cite{sorrentino2016complete}. However, we note that such states still satisfy the constraints of our problem, as they admit the target symmetries (as well as possibly additional symmetries not included in the target set). {In Sec. \ref{sec:numericalresults} we will show numerical examples of scenarios where either coarser cluster states exist or they do not, in particular focusing on a control targeting a number of clusters comprising a subset of the nodes of the entire network.}

{An important consideration in order to assess whether our approach leads to the emergence of the desired cluster state is the coexistence of multiple attractors. In fact, two scenarios may appear when comparing the stability of the target CS manifold with that of any other coarser cluster state existing in the network or of GS.  In the first scenario, the GS state is unstable and the CS state is locally stable, which is typically the case for the coupling in a suitable interval of values \cite{pecora2013symmetries,sorrentino2016complete}. In the second case another coarser CS state (or the GS state) and the CS state are both locally stable. Once established that CS and another coarser CS state (or the GS state) may coexist for the same coupling, there is still the question of the basins of attraction that ultimately will determine towards which of the two attractors the system trajectory will converge. This a fundamental question whenever there are two or more attractors for a dynamical system, but predicting the basins of attraction cannot be answered in general for all oscillators, without more specific knowledge of the oscillators themselves.}

{A related question is how to impose strict control requirements such as that two (or more) clusters are not allowed to converge to the same trajectory. While addressing this problem is beyond the scope of this paper, we can think of a few ways to enforce a specific CS pattern and prevent the merging of clusters. A possibility is to extend the approach presented in \cite{belykh2001cluster} for an array configuration to define general transformations that, when applied to the network, preserve the symmetries but destabilize the GS manifold. Notice however that, in the above mentioned example of the array configuration, the method yields a connectivity which cannot be anymore modeled by a Laplacian and involves many changes in the original structure of interactions. Another possibility is to apply a perturbation to a suitable subset of the network nodes to move the trajectory from the basin of attraction of one attractor to the other, extending the idea applied, via pinning, to the control of chimera states \cite{gambuzza2016pinning}.}

\subsection{{Extension to networks with non-Laplacian coupling}}

{In Eqs. (\ref{eq:eq1Lapl}) interactions among the network nodes are modeled via Laplacian coupling. Here, we consider a different model described by the following equations:}

\begin{equation}
\label{eq:eq1Adj}
\dot{\mathbf{x}}_i=\mathbf{f}(\mathbf{x}_i)+\sigma\sum_{j=1}^N\mathrm{A}_{ij} \mathbf{h}(\mathbf{x}_j) ,
\end{equation}

\noindent {where $A_{ij}$ are the coefficients of the symmetric adjacency matrix. In this model, interactions are no longer diffusive and invariance of the GS manifold is generally not guaranteed, as it is for the case of Laplacian coupling. Eq. (\ref{eq:eq1Adj}) may be of general interest in many different fields, as for instance in the investigation of networks of bursting neurons where relatively large clusters may emerge, representing an instance of cluster synchronization at a mesoscale network level \cite{belykh2011mesoscale}.}

{System (\ref{eq:eq1Adj}) can be dealt with analogously to Eqs. (\ref{eq:eq1Lapl}). In fact, as symmetries of the Laplacian matrix are the same as those of the adjacency matrix \cite{pecora2013symmetries}, the methods presented in Secs. \ref{sec:sylvester} and \ref{sec:optmethods} can be directly applied to Eqs. (\ref{eq:eq1Adj}) as well. }

{The problem of the stability of the CS for Eqs. (\ref{eq:eq1Adj}) can be similarly tackled, by leveraging the results discussed in \cite{pecora2014cluster}. It is here important to note that, as GS is not always guaranteed to be a feasible solution for (\ref{eq:eq1Adj}), enforcing a given CS pattern may be easier for the adjacency matrix case than for the Laplacian case.}

\section{Numerical results}
\label{sec:numericalresults}

In this section, first we consider two illustrative small-size networks and discuss in details the network changes as well as the synchronous pattern obtained. We move then to larger structures where we show the suitability of the proposed optimization-based solutions across several network models.

\subsection{Illustrative small-size networks}

{We now prove validity of our methods in networks of chaotic R\"ossler oscillators. This choice is purely exemplificative as the approach applies to arbitrary, periodic or chaotic dynamics, including for instance the Lorenz system.} For the chaotic R\"ossler oscillator $\mathbf{x}_i = [ x_{i,1}, x_{i,2},  x_{i,3} ]^\mathrm{T}$, and

\begin{equation}
\mathbf{f}(\mathbf{x}_i)=\left [ \begin{array}{c} -x_{i,2}-x_{i,3}\\
x_{i,1}+a_R x_{i,2}\\
b_R+x_{i,3}(x_{i,1}-c_R)  \end{array} \right] \ ,
\end{equation}

\noindent where $a_R$, $b_R$ and $c_R$ are system parameters (in our applications, we used $a_R=b_R=0.2$, $c_R=9$). The coupling function has been fixed as:

\begin{equation}
\label{eq:couplingH}
\mathbf{h}(\mathbf{x}_i)=\left [ \begin{array}{c}0\\
x_{i,2}\\
0  \end{array} \right] \ .
\end{equation}

\noindent {such that the system is in class II. Note that our analysis focuses on the first threshold for synchronization, which is a feature common to class II and class III systems, so that our considerations are completely general and apply to both cases.}

CS is monitored by the error  $e_{V_l}(t)= \sqrt { \sum\limits_{i,j\in V_l} \parallel \mathbf{x}_i(t) - \mathbf{x}_j(t) \parallel ^2}$ for each $V_l$ with $l=1,\ldots,M$. $e_{V_l}(t)=0$ if all units of cluster $V_l$ are synchronized. When a cluster only contains two units, a shorter notation is used: $e_{k,l}(t)= \parallel \mathbf{x}_k(t) - \mathbf{x}_{l}(t) \parallel$. On the other hand, the error $e(t)=\sqrt { \sum_{\substack{i,j}} \parallel \mathbf{x}_i(t) - \mathbf{x}_j(t) \parallel ^2}$ is used to monitor GS.

We consider the graph with $N=12$ nodes shown in Fig.~\ref{fig:Ex1A}, which does not displays any symmetry. We here want to impose a target symmetry defined by the permutations $1\leftrightarrow 2$, $7\leftrightarrow 8$, $9\leftrightarrow 10$ and $11\leftrightarrow 12$, so that the emerging CS state will then have units 1 and 2 (blue nodes) in the same cluster, while units 7 and 8 (brown nodes), 9 and 10 (green nodes), and 11 and 12 (pink nodes) will form other three clusters. Here, $Q=1$, so that $\mathcal{H}$ contains only the target symmetry $\mathrm{R}_1$ and the identity.

In this case $Q=1$ and the permutation matrix associated to the desired clusters is:

\begin{equation}
\mathrm{R}_1=\left(\begin{array}{llllllllllll}
0 & 1 & 0 & 0 & 0 & 0 & 0 & 0 & 0 & 0 & 0 & 0\\
1 & 0 & 0 & 0 & 0 & 0 & 0 & 0 & 0 & 0 & 0 & 0\\
0 & 0 & 1 & 0 & 0 & 0 & 0 & 0 & 0 & 0 & 0 & 0\\
0 & 0 & 0 & 1 & 0 & 0 & 0 & 0 & 0 & 0 & 0 & 0\\
0 & 0 & 0 & 0 & 1 & 0 & 0 & 0 & 0 & 0 & 0 & 0\\
0 & 0 & 0 & 0 & 0 & 1 & 0 & 0 & 0 & 0 & 0 & 0\\
0 & 0 & 0 & 0 & 0 & 0 & 0 & 1 & 0 & 0 & 0 & 0\\
0 & 0 & 0 & 0 & 0 & 0 & 1 & 0 & 0 & 0 & 0 & 0\\
0 & 0 & 0 & 0 & 0 & 0 & 0 & 0 & 0 & 1 & 0 & 0\\
0 & 0 & 0 & 0 & 0 & 0 & 0 & 0 & 1 & 0 & 0 & 0\\
0 & 0 & 0 & 0 & 0 & 0 & 0 & 0 & 0 & 0 & 0 & 1\\
0 & 0 & 0 & 0 & 0 & 0 & 0 & 0 & 0 & 0 & 1 & 0\\
\end{array}\right) \ .
\end{equation}

\begin{figure}
\begin{center}
\subfigure[]{\includegraphics[width=0.24\textwidth]{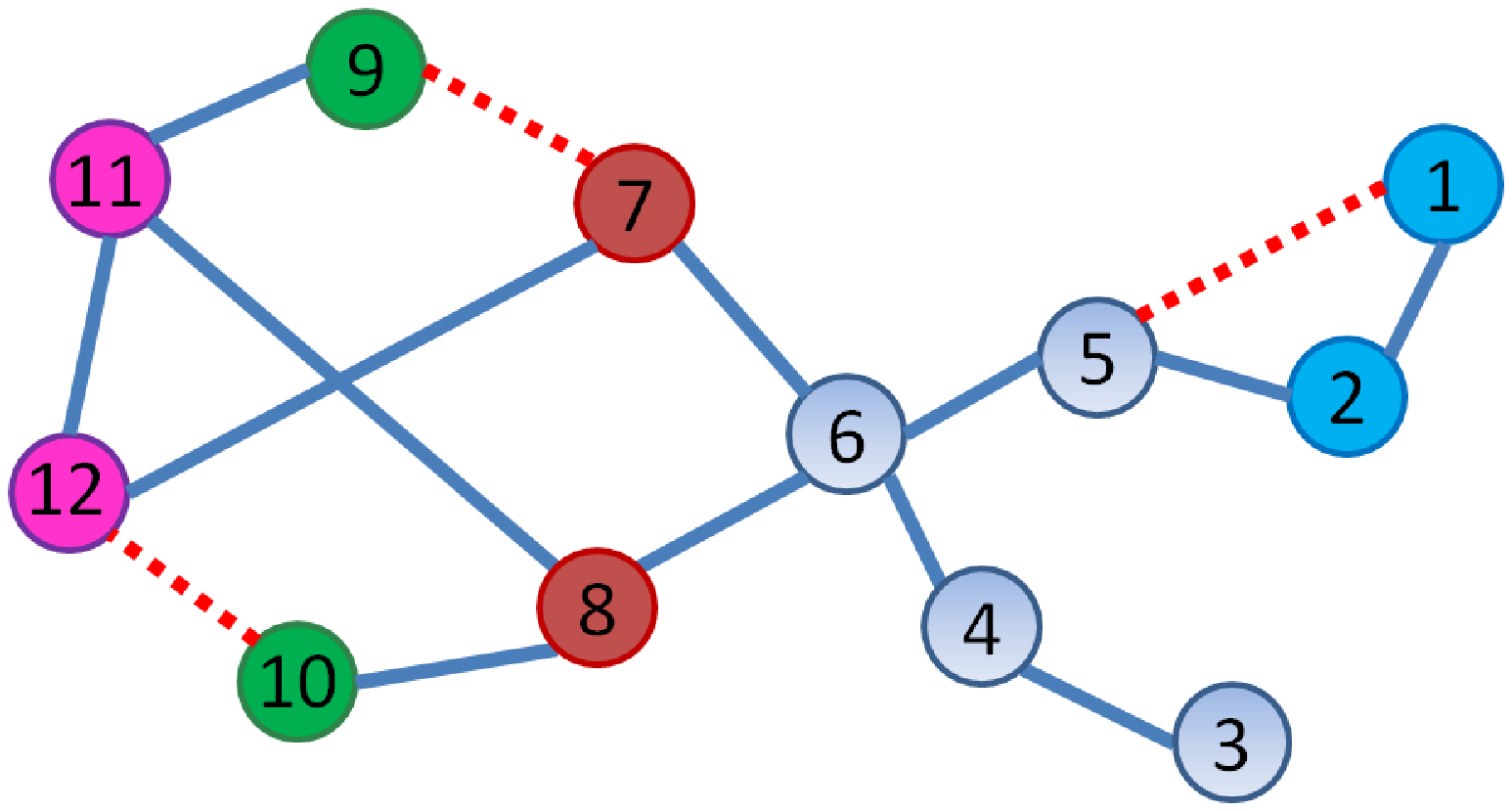}\label{fig:Ex1secondocriterioA}}
\subfigure[]{\includegraphics[width=0.23\textwidth]{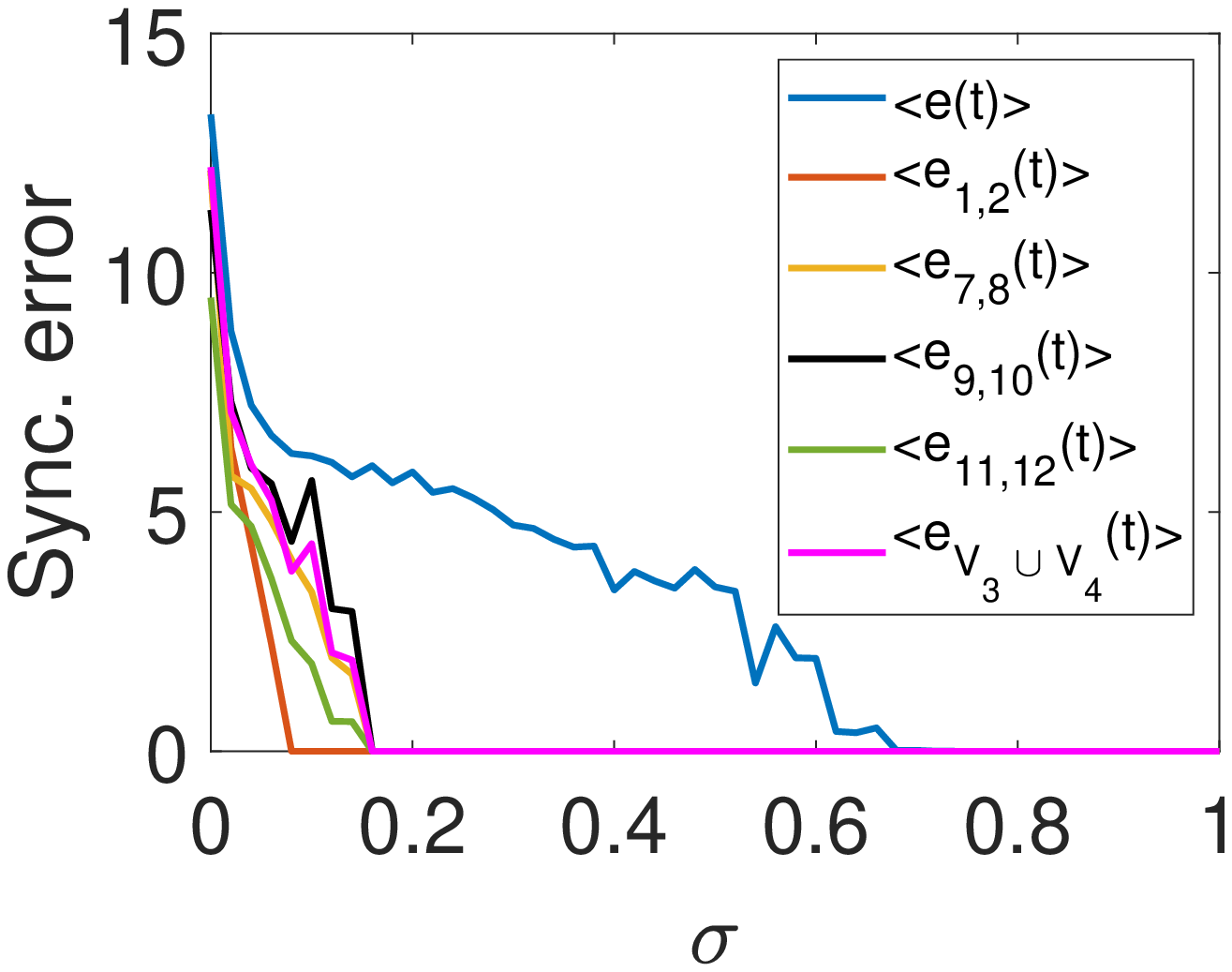}\label{fig:Ex1secondocriterioB}}
\caption{\label{fig:Ex1secondocriterio} Controlling with unweighted graphs. (a) The resulting controlled topology: the three added links are marked in red (dashed) lines; (b) synchronization error vs. $\sigma$ for the relevant clusters and the whole network.}
\end{center}
\end{figure}

\begin{figure}
\begin{center}
\subfigure[]{\includegraphics[width=0.16\textwidth]{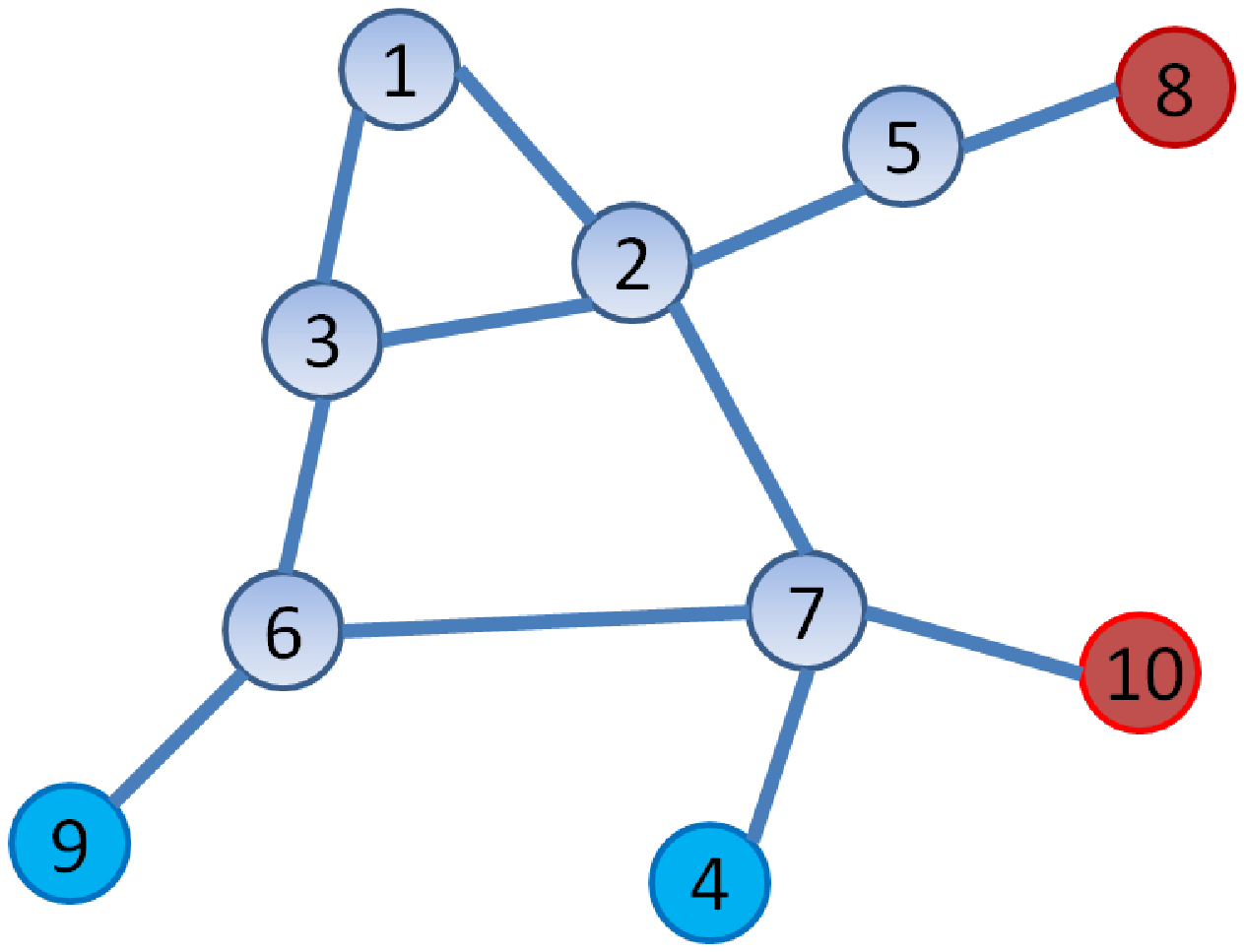}\label{fig:Ex2A}}
\subfigure[]{\includegraphics[width=0.16\textwidth]{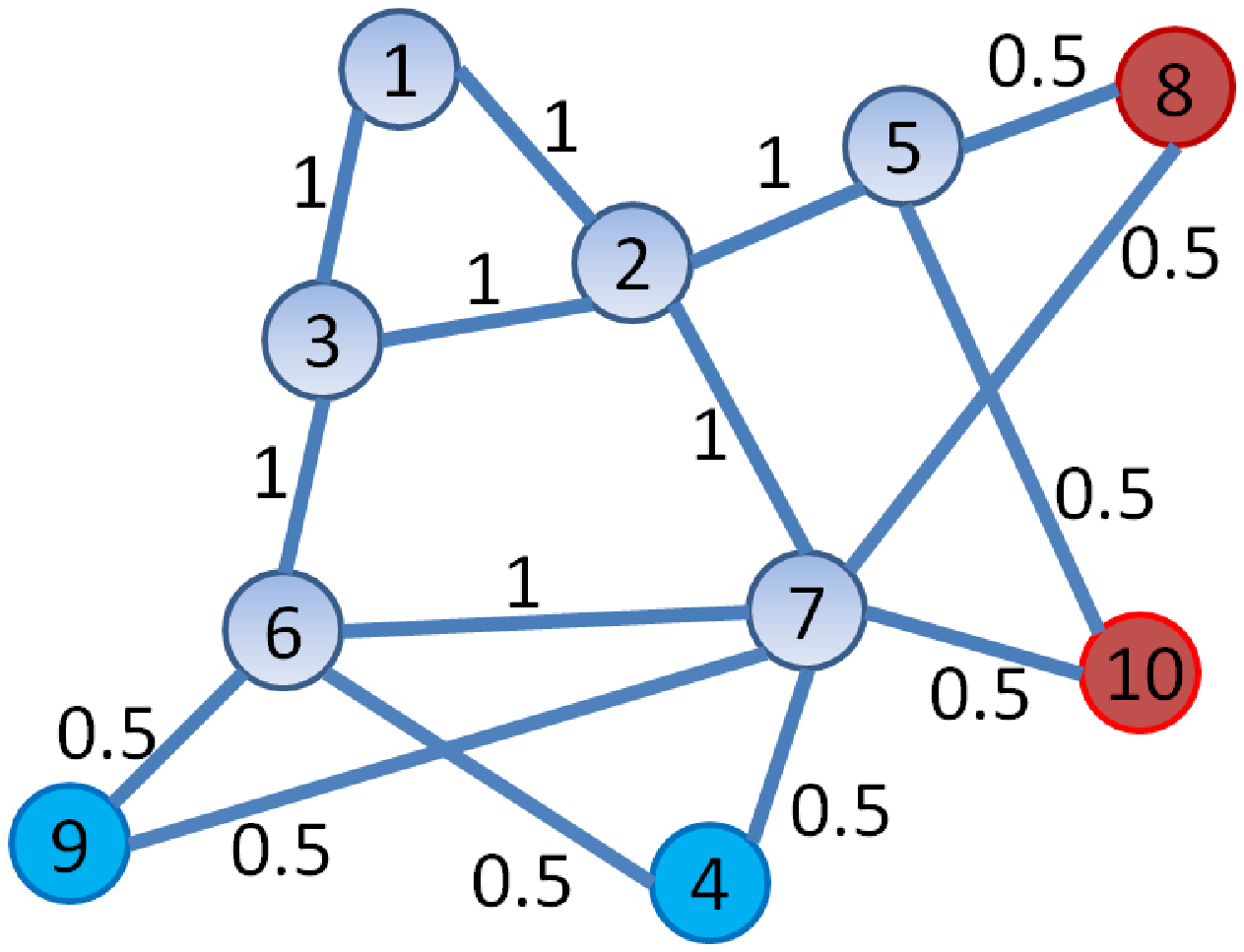}\label{fig:Ex2B}}
\subfigure[]{\includegraphics[width=0.16\textwidth]{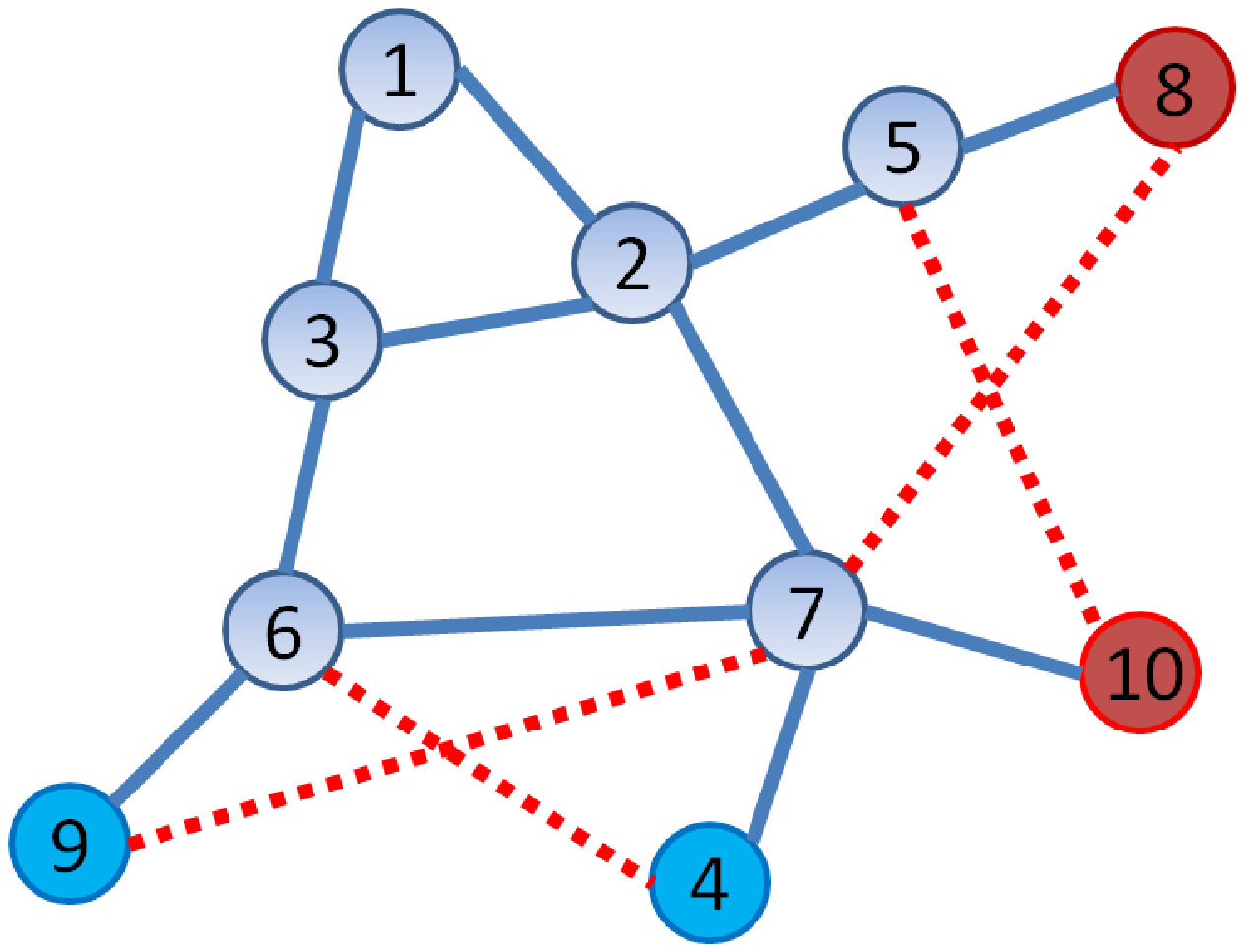}\label{fig:Ex2C}}
\caption{\label{fig:Ex2} Controlling clusters with $Q=2$. (a) Sketch of the pristine network; (b) weighted controlled network: notice cluster $V_1$ (involving the blue nodes 4 and 9) and cluster $V_2$ (involving the brown nodes 8 and 10); (c) unweighted controlled network with added links marked in red (dashed) lines.}
\end{center}
\end{figure}

The network obtained by the MPI method is depicted in Fig.~\ref{fig:Ex1B}: one can notice that three links have been added and the weights of several others are changed. The new network displays the following orbit partition: $\{1,2\}$, $\{7,8\}$, $\{9,10\}$, $\{11,12\}$, $\{3\}$, $\{4\}$, $\{5\}$, $\{6\}$, thus a CS state is generated with $M=8$ clusters: $V_1=\{1,2\}$, $V_2=\{7,8\}$, $V_3=\{9,10\}$, $V_4=\{11,12\}$, $V_5=\{3\}$, $V_6=\{4\}$, $V_7=\{5\}$, $V_8=\{6\}$.
The variational equations for the transverse modes (obtained
by application of the matrix $\mathrm{T}$ corresponding to the irreducible representation of  $\mathcal{H}$), can be written as

\begin{equation}
\label{eq:Stability1pesata}
\begin{array}{lll}
\delta \mathbf{\dot{y}}_9 & = & \left [ \mathbf{Df}(\mathbf{s}_2) - 2.5 \sigma\mathbf{Dh}(\mathbf{s}_2)\right] \delta \mathbf{y}_9 -  \sigma\mathbf{Dh}(\mathbf{s}_4)\delta \mathbf{y}_{10} \\
& & +0.5 \sigma\mathbf{Dh}(\mathbf{s}_3)\delta \mathbf{y}_{11} \  , \\
\delta \mathbf{\dot{y}}_{10} & = & \left [ \mathbf{Df}(\mathbf{s}_4) - 3.5 \sigma\mathbf{Dh}(\mathbf{s}_4)\right] \delta \mathbf{y}_{10} -  \sigma\mathbf{Dh}(\mathbf{s}_2)\delta \mathbf{y}_{9} \\
& & +0.5 \sigma\mathbf{Dh}(\mathbf{s}_3)\delta \mathbf{y}_{11} \ , \\
 \delta \mathbf{\dot{y}}_{11} & = & \left [ \mathbf{Df}(\mathbf{s}_3) - \sigma\mathbf{Dh}(\mathbf{s}_3)\right] \delta \mathbf{y}_{11} + 0.5 \sigma\mathbf{Dh}(\mathbf{s}_2)\delta \mathbf{y}_{9} \\
& & +0.5 \sigma\mathbf{Dh}(\mathbf{s}_4)\delta \mathbf{y}_{10} \ , \\
\delta \mathbf{\dot{y}}_{12} & = & \left [ \mathbf{Df}(\mathbf{s}_1) - 2.5 \sigma\mathbf{Dh}(\mathbf{s}_1)\right] \delta \mathbf{y}_{12} \ .
\end{array}
\end{equation}

Note that the set of equations in ${\delta \mathbf{y}}_9$, ${\delta \mathbf{y}}_{10}$ and ${\delta \mathbf{y}}_{11}$ correspond to three intertwined clusters, $V_2$, $V_3$, and $V_4$, while the stability of the cluster $V_1$ is independent from them.

From these latter equations, the maximum Lyapunov exponent is calculated as a function of $\sigma$. The result is reported in Fig.~\ref{fig:Ex1C}, where one can see that the CS state is stable for $\sigma>0.2$. In the same figure, the maximum Lyapunov exponent for GS is also shown, which indicates that GS becomes stable for larger values of $\sigma$. The evolution of the errors for GS and for the four non-trivial clusters $V_1,\ldots,V_4$ is illustrated in Figs.~\ref{fig:Ex1D}-\ref{fig:Ex1H}. In our simulations, the pristine network of Fig.~\ref{fig:Ex1A} is initially (for $t<t_c=500$) used for coupling the units, while for $t>t_c=500$ the controlled network of Fig.~\ref{fig:Ex1B} is enforced. The results confirm that, at this value of $\sigma$, CS is stable in the new graph but not in the original one, and that GS is not observed.

The solution obtained by the CPO binary optimization problem is sketched in Fig.~\ref{fig:Ex1secondocriterioA}, and one sees that three links (red dashed lines) have been added. {Fig.~\ref{fig:Ex1secondocriterioB} illustrates the system behavior at different coupling values, by showing the average value of the synchronization error for the relevant clusters and for the whole network vs. $\sigma$ (simulations are carried out over a time period $T=2000$ and the average is over the last T/2 period). In the interval $\sigma \in [0.16, 0.7]$ the system displays a CS state that satisfies the target set of symmetries. As it can be observed, this CS state is coarser than the one obtained by the MPI method in Fig.~\ref{fig:Ex1B}, as here 
the two clusters $V_3=\{9,10\}$ and $V_4=\{11,12\}$ merged together (in this interval the error $<e_{V_3 \cup V_4}(t)>$ is also zero). Also note that the cluster $V_1=\{1,2\}$ becomes stable for smaller values of $\sigma$, i.e., $\sigma\geq 0.08$. GS, instead, becomes stable for $\sigma \simeq 0.7$. It is relevant that the coarser CS appears in the unweighted controlled network, but not in the controlled weighted structure, where the onset of this state is prevented by differences in the link weights.}

{Although preserving sparsity is more relevant in larger networks, the lasso method can still be applied to the case study of Fig. \ref{fig:Ex1A}. The method finds that three links need to be changed; in particular, two links, (1,5) and (7,9), are added, and a link, (9,11), is removed. The result is another network realization that admits the desired symmetries with a minimal number of changes (three).}

As a second application, we consider the network of Fig.~\ref{fig:Ex2A}, which has a unique symmetry between nodes 4 and 10.
We consider $Q=2$, and two matrices $\mathrm{R}_1$ and $\mathrm{R}_2$ as target symmetries such that
the units 4 and 9 (blue nodes) and the units 8 and 10 (brown nodes) will cluster together in the controlled CS state:

\begin{equation}
\mathrm{R}_1=\left(\begin{array}{llllllllll}
1 & 0 & 0 & 0 & 0 & 0 & 0 & 0 & 0 & 0\\
0 & 1 & 0 & 0 & 0 & 0 & 0 & 0 & 0 & 0\\
0 & 0 & 1 & 0 & 0 & 0 & 0 & 0 & 0 & 0\\
0 & 0 & 0 & 0 & 0 & 0 & 0 & 0 & 1 & 0\\
0 & 0 & 0 & 0 & 1 & 0 & 0 & 0 & 0 & 0\\
0 & 0 & 0 & 0 & 0 & 1 & 0 & 0 & 0 & 0\\
0 & 0 & 0 & 0 & 0 & 0 & 1 & 0 & 0 & 0\\
0 & 0 & 0 & 0 & 0 & 0 & 0 & 1 & 0 & 0\\
0 & 0 & 0 & 1 & 0 & 0 & 0 & 0 & 0 & 0\\
0 & 0 & 0 & 0 & 0 & 0 & 0 & 0 & 0 & 1
\end{array}\right)
\end{equation}

\noindent and

\begin{equation}
\mathrm{R}_2=\left(\begin{array}{llllllllll}
1 & 0 & 0 & 0 & 0 & 0 & 0 & 0 & 0 & 0\\
0 & 1 & 0 & 0 & 0 & 0 & 0 & 0 & 0 & 0\\
0 & 0 & 1 & 0 & 0 & 0 & 0 & 0 & 0 & 0\\
0 & 0 & 0 & 1 & 0 & 0 & 0 & 0 & 0 & 0\\
0 & 0 & 0 & 0 & 1 & 0 & 0 & 0 & 0 & 0\\
0 & 0 & 0 & 0 & 0 & 1 & 0 & 0 & 0 & 0\\
0 & 0 & 0 & 0 & 0 & 0 & 1 & 0 & 0 & 0\\
0 & 0 & 0 & 0 & 0 & 0 & 0 & 0 & 0 & 1\\
0 & 0 & 0 & 0 & 0 & 0 & 0 & 0 & 1 & 0\\
0 & 0 & 0 & 0 & 0 & 0 & 0 & 1 & 0 & 0
\end{array}\right) \ .
\end{equation}

The solution obtained with the MPI method (a weighted network) is shown in Fig.~\ref{fig:Ex2B}, and the solution obtained through CPO binary linear programming (an unweighted network) in Fig.~\ref{fig:Ex2C}. In both cases, the controlled graph induces the same neighborhood in the pairs of nodes to be synchronized. The final configuration has only two orbit partitions, $\{4,9\}$, and $\{8,10\}$, indicating that the symmetry existing in the original network has been destroyed by the addition of the new links. The associated CS state has two non-trivial clusters $V_1=\{4,9\}$, $V_1=\{8,10\}$ and six singletons formed by the remaining nodes. The pattern is stable for $\sigma > 0.2$ in the network of Fig.~\ref{fig:Ex2B} and for $\sigma > 0.1 $ in the network of Fig.~\ref{fig:Ex2C}.

{Using the lasso method for the network of Fig. \ref{fig:Ex2A}, one finds that four link changes are required. In particular, 
the links  (4,6) and (7,8) need to be added, and the links (4,7) and (5,8) need to be removed. Despite the resulting network is different from that found by the CPO method we observe that the number of changes is the same.}

\subsection{Large Networks}

The above examples refer to two illustrative small-size networks, but the methods here presented are general and can be applied to large networks as well. In particular, all the three methods of Sec. \ref{sec:optmethods} may be suitably applied using sparse matrices, which is the typical case of real-world networks having a number of links that does not scale with $N^2$, $N$ being the number of the elementary constituents (the network nodes). We apply the three methods to diverse scenarios, in order to illustrate their performance when applied to different topologies and the scaling of their computational demand with $N$ and with the number of generators in the symmetry group. In each scenario, we monitor the $L_2$-norm of the obtained solution, i.e., $\|\Delta A\|_2$, and the percentage number of modified links, i.e., $n_{L,\%}=\frac{n_L}{N(N-1)}\cdot 100$, where $n_L$ is the number of links that are changed in the network after the application of the control.

\begin{figure}[h]
\subfigure[]{\includegraphics[width=0.24\textwidth]{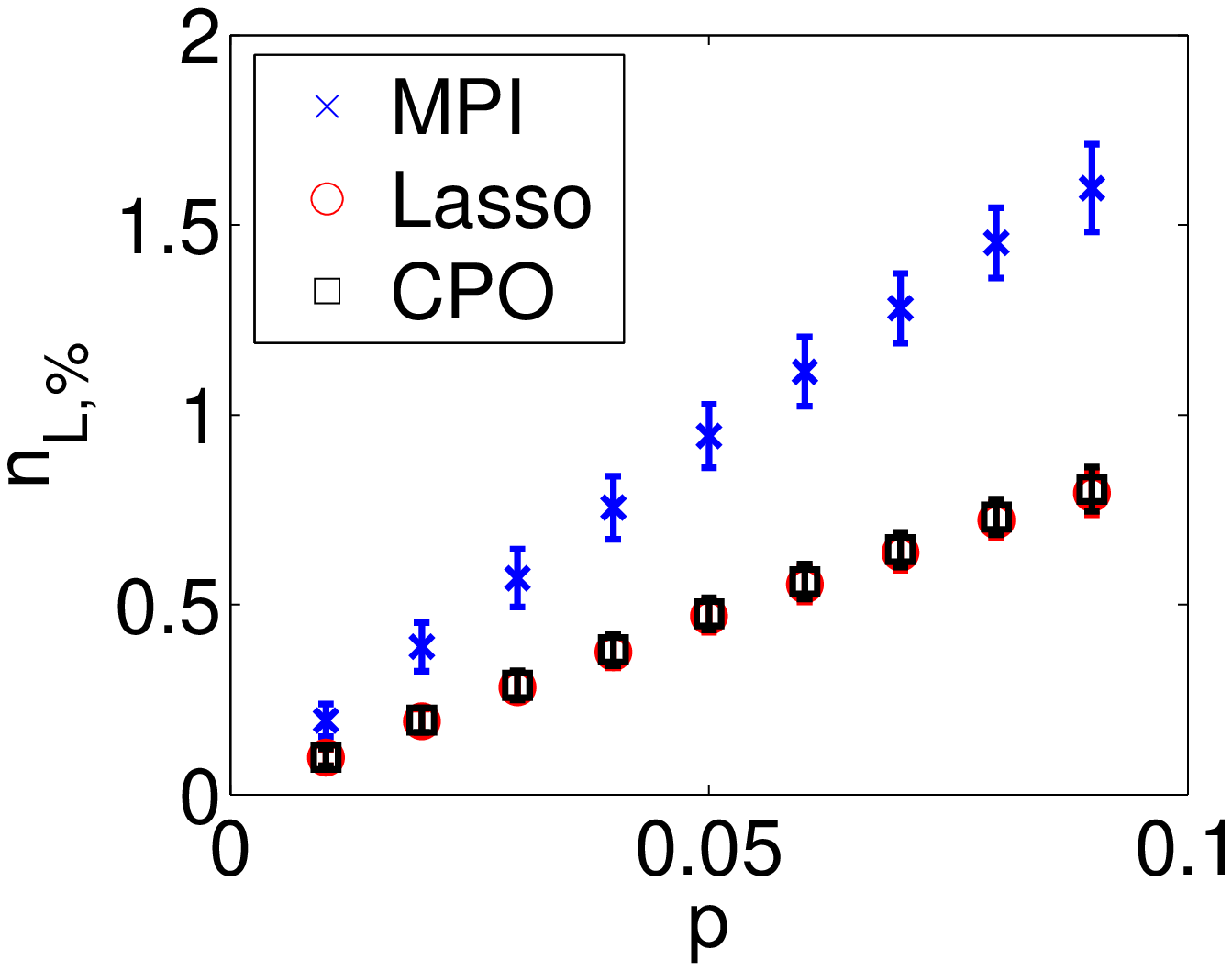}}
\subfigure[]{\includegraphics[width=0.24\textwidth]{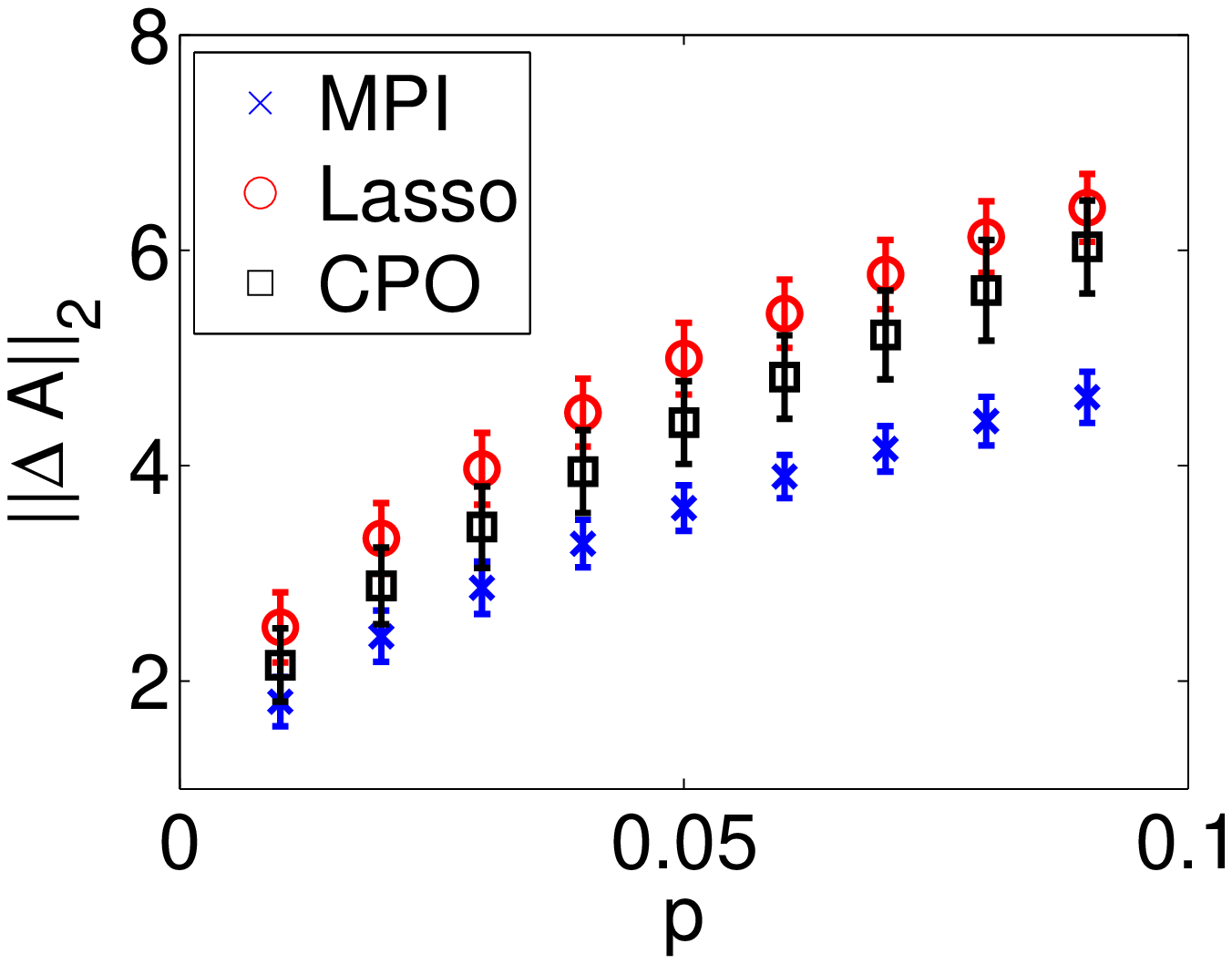}}
\caption{\label{fig:NewResultsER} Inducing symmetries in ER networks: (a) $n_{L,\%}$; (b) $\|\Delta A\|_2$. Networks are parametrized with $p$, the link probability. Increasing values of $p$ thus represent networks with increasing connectivity. All the networks have $N=200$ nodes. The group symmetry is characterized by $Q=2$, with one target cluster having two random nodes and a second one having three random nodes. Results are averaged over 100 networks, with error bars indicating the standard deviation from the average.}
\end{figure}

\begin{figure}[h]
\subfigure[]{\includegraphics[width=0.24\textwidth]{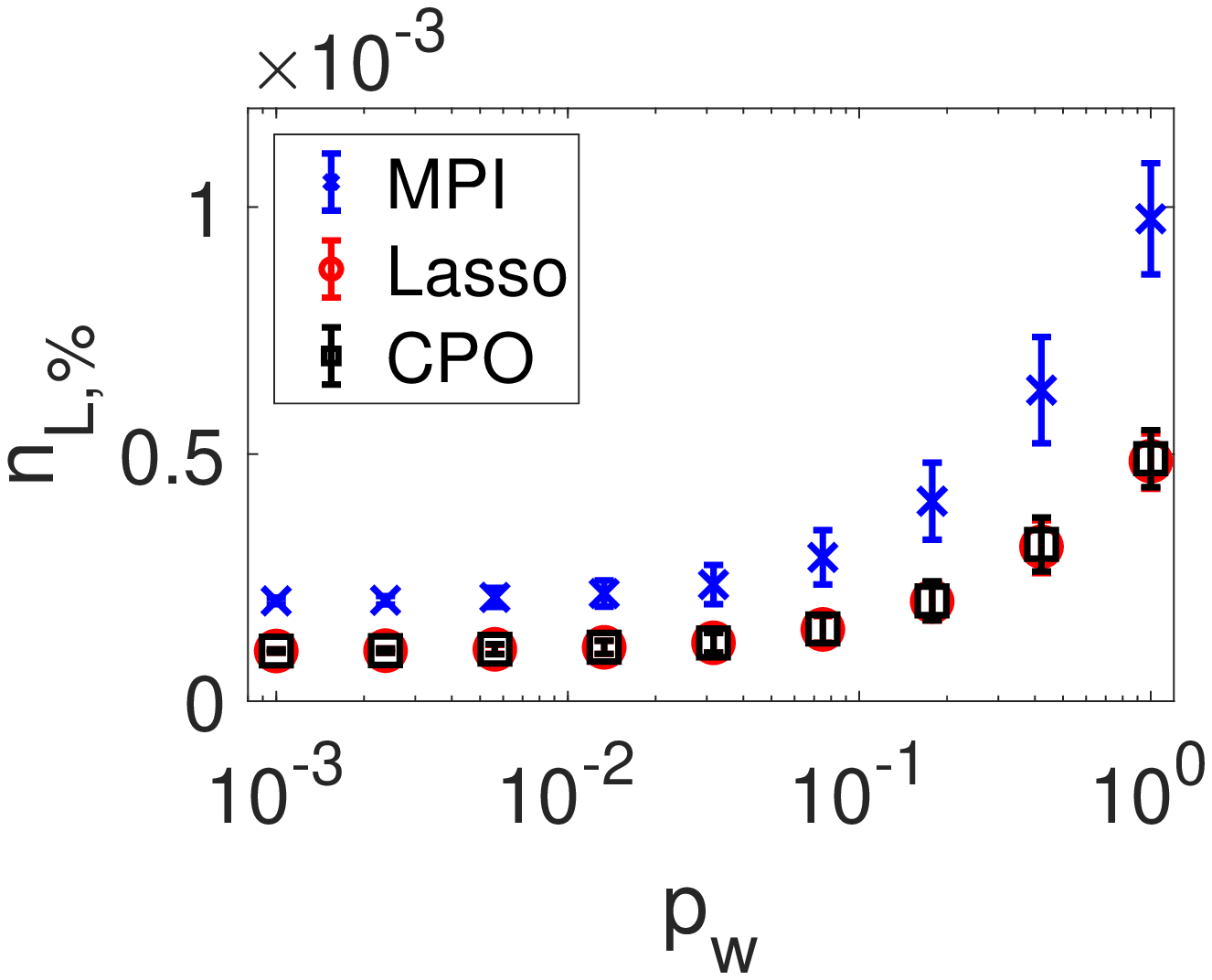}}
\subfigure[]{\includegraphics[width=0.24\textwidth]{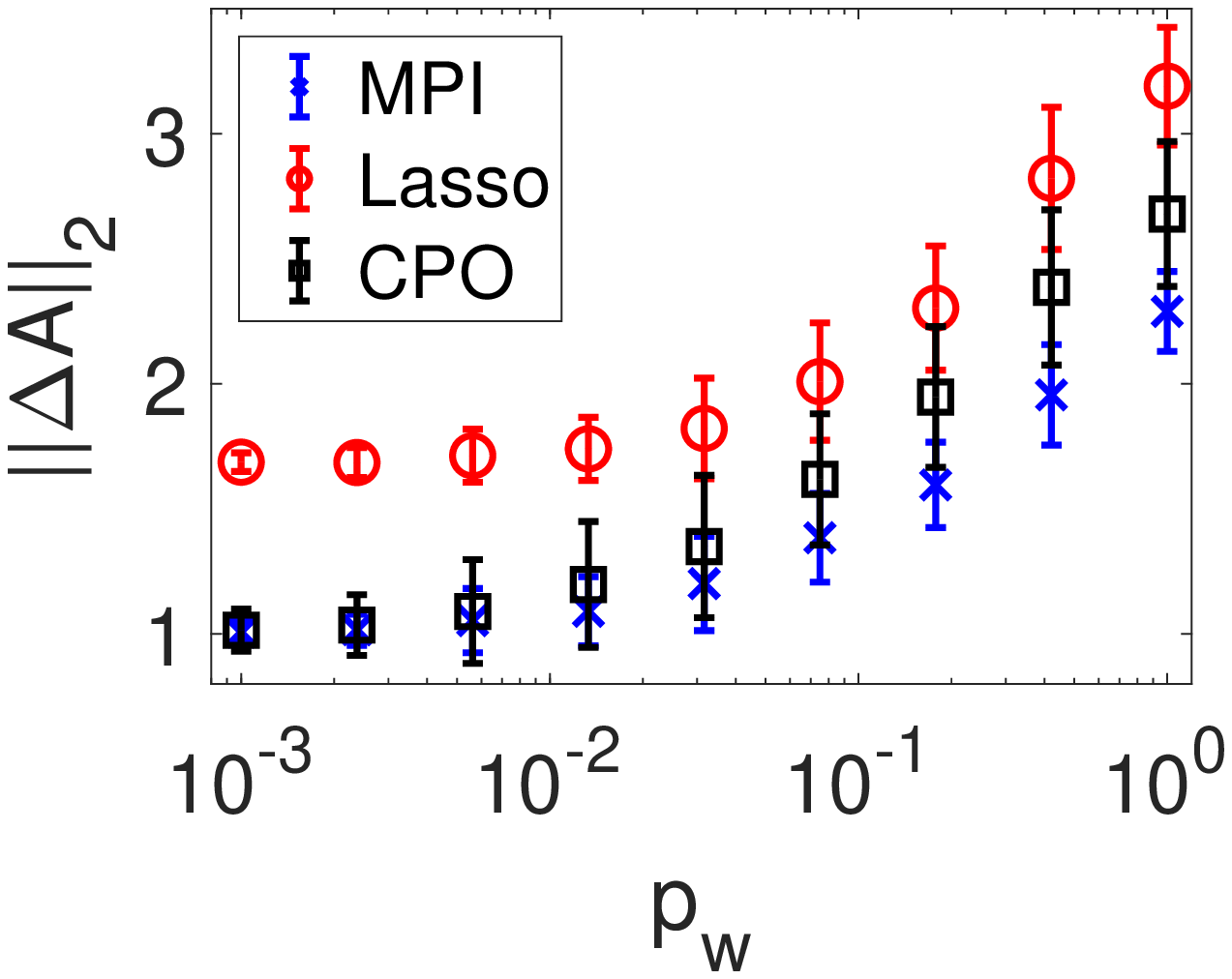}}
\caption{\label{fig:NewResultsWS} Inducing symmetries in WS networks: (a) $n_{L,\%}$; (b) $\|\Delta A\|_2$. Networks are built by considering a 2-neighbor ring of each node, and rewiring the links with probability $p_w$. All the networks have $N=200$ nodes. The group symmetry is characterized by $Q=2$, with one target cluster having two random nodes and a second one having three random nodes. Results are averaged over 100 networks, with error bars indicating the standard deviation from the average.}
\end{figure}

\begin{figure}[h]
\subfigure[]{\includegraphics[width=0.24\textwidth]{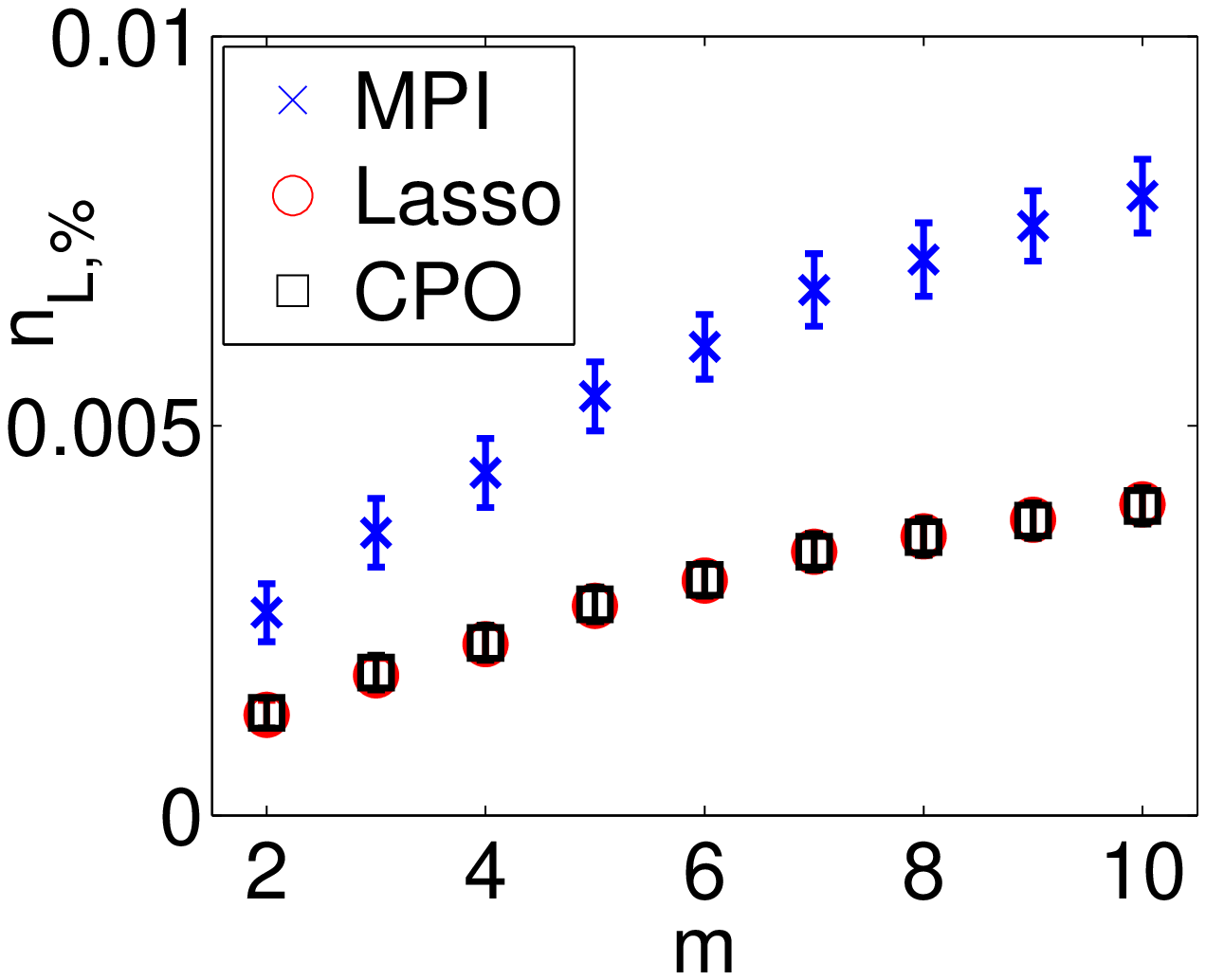}}
\subfigure[]{\includegraphics[width=0.24\textwidth]{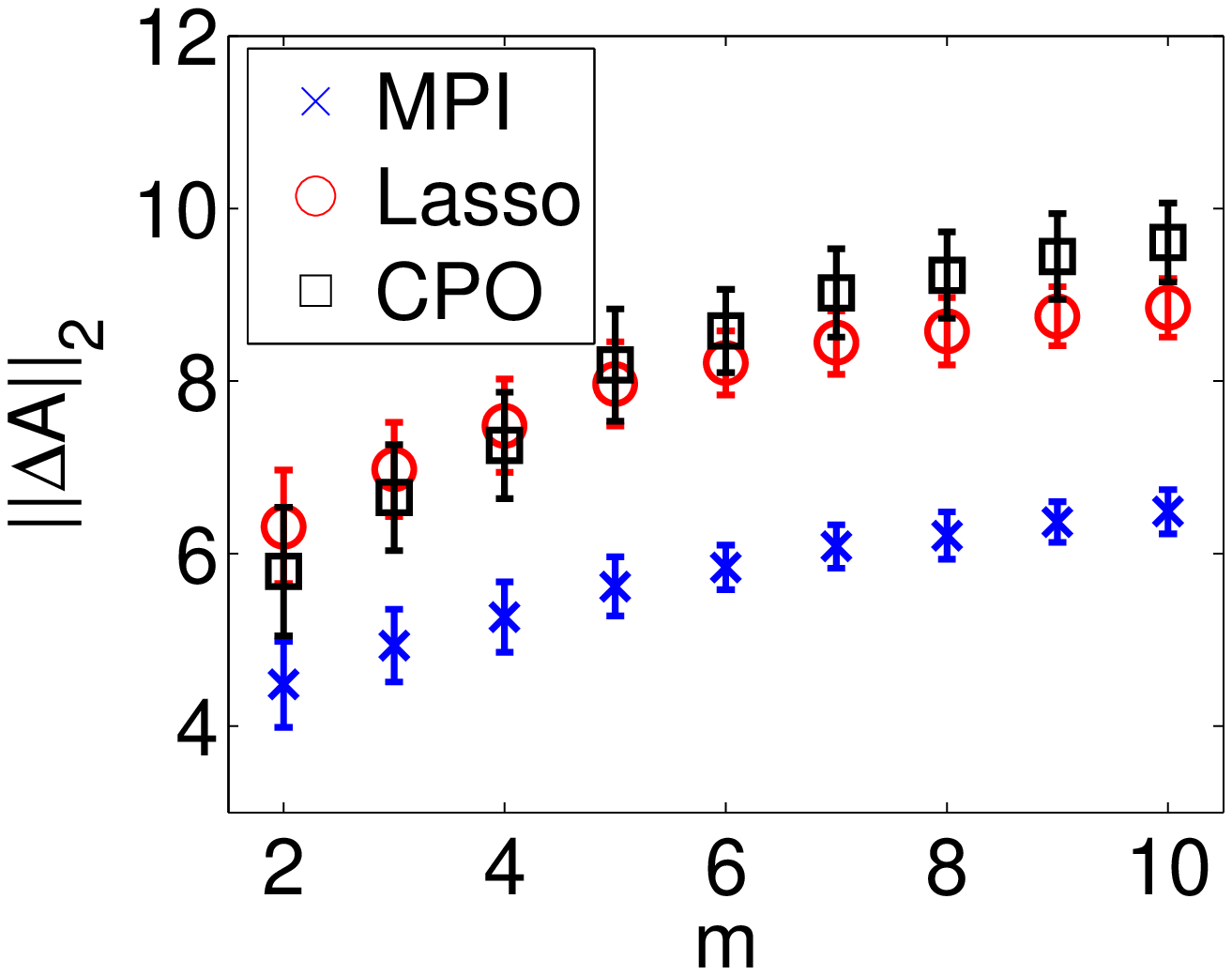}}
\caption{\label{fig:NewResultsBA} Inducing symmetries in BA networks: (a) $n_{L,\%}$; (b) $\|\Delta A\|_2$. Networks are parametrized with $m$, i.e., the number of neighbors added at each step in the growing process generating the structure. All the networks have $N=200$ nodes. The group symmetry is characterized by $Q=2$, with one target cluster having two random nodes and a second one having three random nodes. Results are averaged over 100 networks, with error bars indicating the standard deviation from the average.}
\end{figure}

We begin  with synthetic examples of random complex networks, such as Erd\H{o}s-R\'enyi (ER) graphs, Watts-Strogats (WS) model of small-world networks, and Barabasi-Albert (BA) model of scale-free networks (the three models are described in detail in Ref.\cite{latora2017} along with the codes for their generation). Figures \ref{fig:NewResultsER}, \ref{fig:NewResultsWS} and \ref{fig:NewResultsBA} illustrate the results for ER, WS and BA networks, respectively. In all  cases, the lasso method provides the solution with the smallest percentage number of modified links, $n_{L,\%}$, while the MPI method provides the solution with the smallest value of $\|\Delta A\|_2$. The CPO method has intermediate performance in terms of $\|\Delta A\|_2$ and values of $n_{L,\%}$, very close to those of the lasso method, but has the advantage of guaranteeing the connectedness of the network. {Notice that for small $p$ the ER graphs are typically not connected, as our method does not require the connectedness of the original network. In such cases, the GS stable is intrinsically unstable.}

\begin{figure}
\subfigure[]{\includegraphics[width=0.24\textwidth]{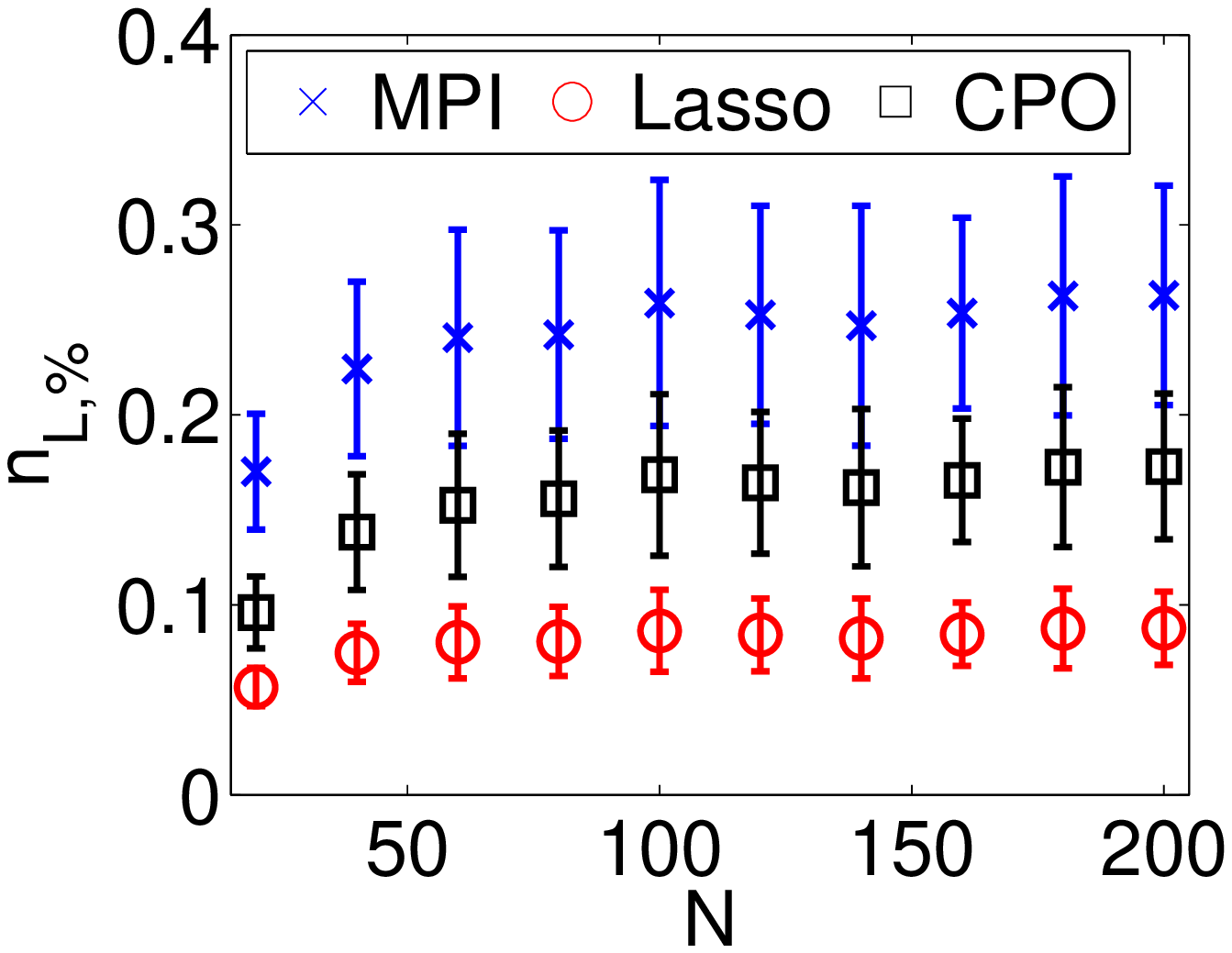}}
\subfigure[]{\includegraphics[width=0.24\textwidth]{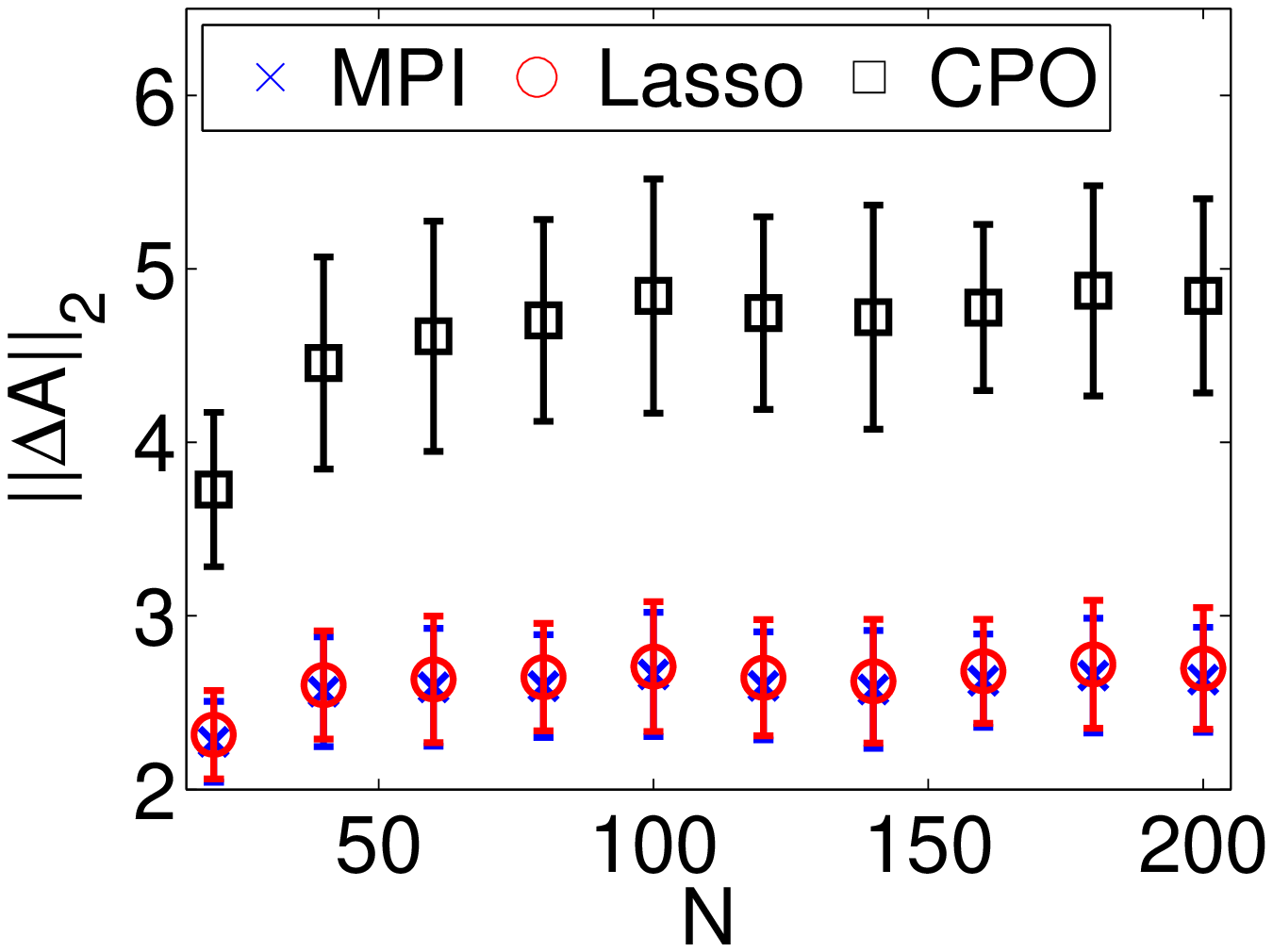}}
\caption{\label{fig:NewResultsERwithN} Scaling with $N$ of the methods to induce symmetries in a network: (a) $n_{L,\%}$; (b) $\|\Delta A\|_2$. Results refer to ER networks with constant average degree $\langle k \rangle=p(N-1)\simeq 6$. The group symmetry is characterized by $Q=2$, with one target cluster having two random nodes and a second one having three random nodes. Results are averaged over 100 networks, with error bars indicating the standard deviation from the average.}
\end{figure}

Next, we analyze the scaling with $N$. In particular, we consider ER networks with different $N$ and constant average degree, i.e., $\langle k \rangle=p(N-1)\simeq 6$. The results are illustrated in Fig. \ref{fig:NewResultsERwithN}, showing that the two parameters weakly depend on $N$, thus demonstrating the applicability of the three methods to general networks. The comparison of the three methods leads to conclusions similar to the previous analysis, with lasso having the smallest percentage number of modified links, $n_{L,\%}$, MPI the smallest value of $\|\Delta A\|_2$, and CPO intermediate performance.

\begin{figure}
\subfigure[]{\includegraphics[width=0.24\textwidth]{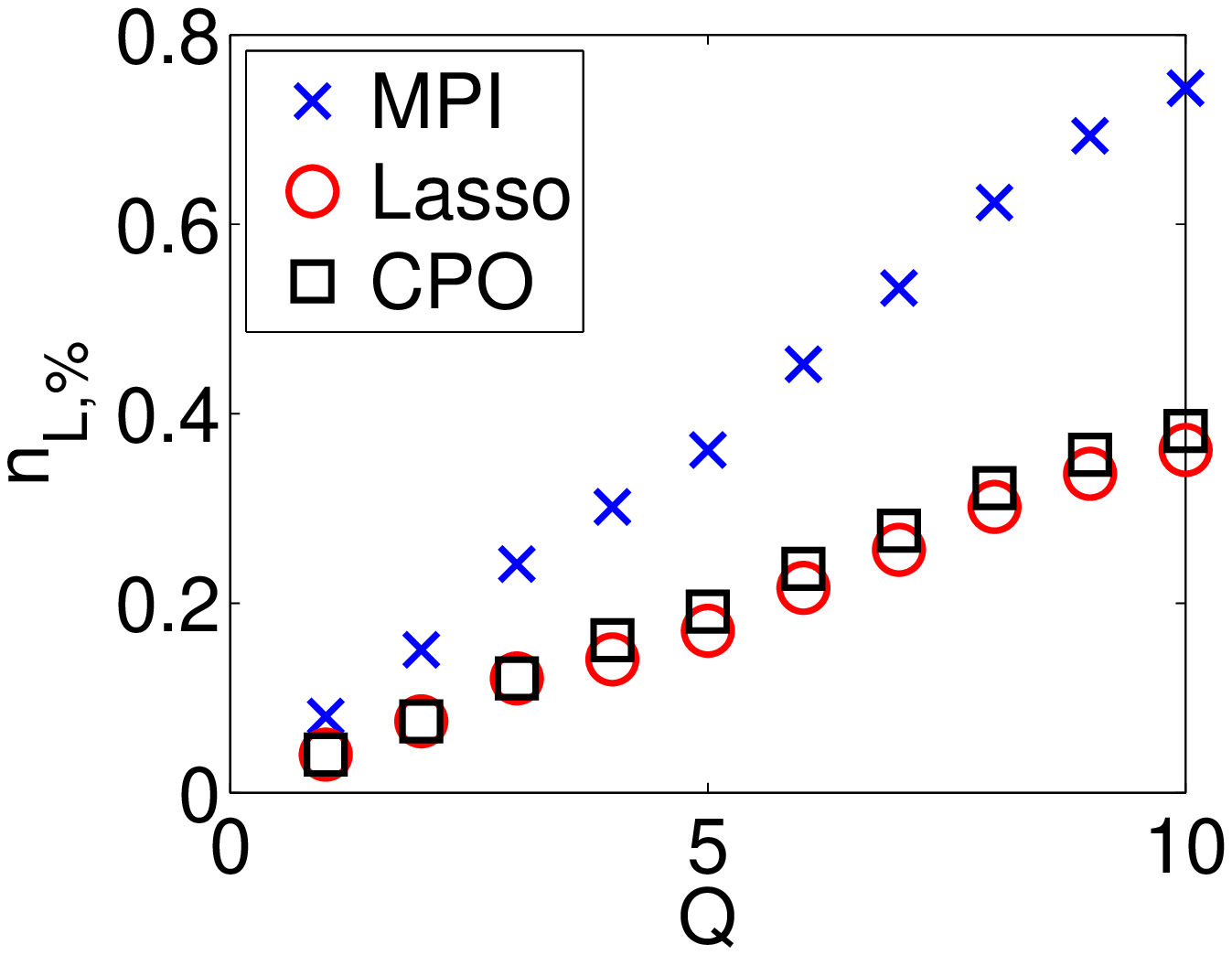}}
\subfigure[]{\includegraphics[width=0.24\textwidth]{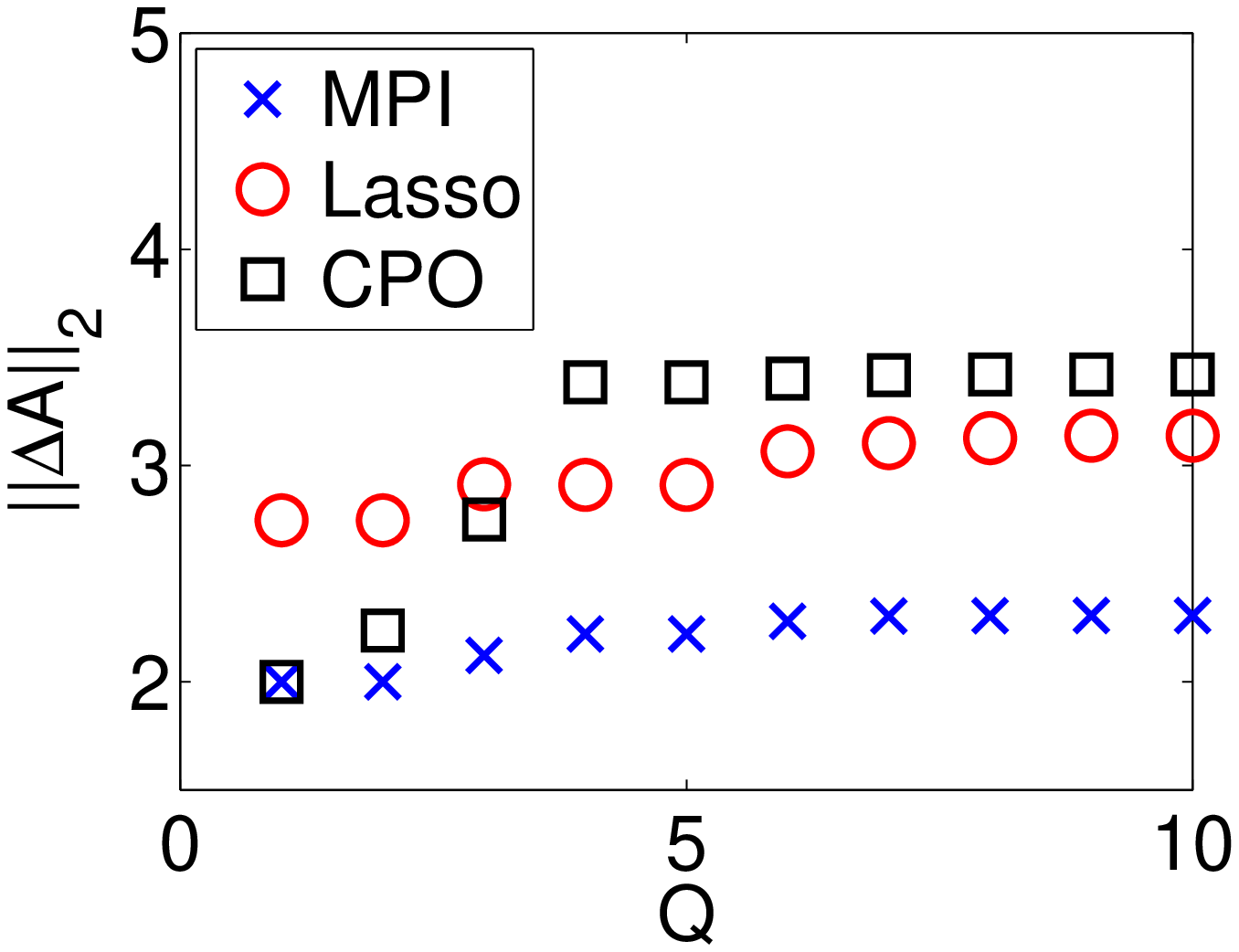}}
\caption{\label{fig:NewResultsERwithQ} Performance of the methods to induce symmetries in a network vs. $Q$: (a) $n_{L,\%}$; (b) $\|\Delta A\|_2$. The network considered here is an ER network with $N=200$ and $p=0.02$.}
\end{figure}

Finally, we analyze the result of the application of the three methods to target symmetries with a growing number of generators $Q$. The results are illustrated in Fig. \ref{fig:NewResultsERwithQ}, which shows that $n_{L,\%}$ scales linearly with $Q$ with slopes different from method to method, while $\|\Delta A\|_2$ slightly increases with $Q$. Even in this case, lasso is the method providing the smallest $n_{L,\%}$, while MPI that providing the smallest $\|\Delta A\|_2$.

{The results presented are intended to show the applicability of our method to arbitrary topologies under different circumstances. It would be interesting to extend the analysis to larger networks or to larger clusters up to the case where every node is assigned to a cluster. A detailed discussion of all these relevant cases, where our approach may be applied, is left for future investigation.}

\section{Applications}

The range of applications of our approach is broad and includes all situations in which one may want to enforce a given set of symmetries in a network, such as cluster synchronization, traffic dynamics on networks, and network evolutionary game theory. Here, we briefly outline two applications where inducing symmetries and controlling the synchronous pattern may be relevant, demonstrating the suitability of our approach to real-world cases. The first example refers to power grids, while the second example to central pattern generators.

\subsection{Power grids}

\begin{figure}
\includegraphics[width=0.5\textwidth]{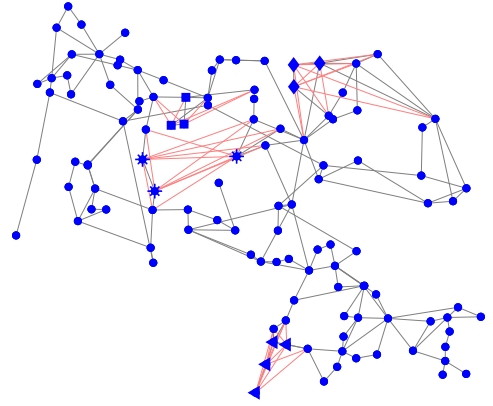}
\caption{\label{fig:PowerGrid} Control of symmetries in the IEEE 118-bus test case, representing a portion of the American Electric Power system in the Midwestern US. The nodes belonging to the four clusters of the symmetry group are represented with filled triangles, diamonds, squares, and asterisks. The links added as result of the application of the CPO method are reported in red.}
\end{figure}

As a first example, we consider here control of symmetries in a power grid network. In particular, we analyze the IEEE 118-bus test case, which represents a portion of the American Electric Power system in the Midwestern US as of December 1962 \cite{database118}. The network consists of $N=118$ nodes (54 are synchronous machines and 64 load stations) and 179 links. Our goal is to cluster synchronize four clusters of nodes, each one composed of 4, 3, 3, and 3 nodes respectively. These nodes are represented with different symbols (filled triangles, diamonds, squares, and asterisks) in Fig. \ref{fig:PowerGrid} (more precisely, according to the labeling in Ref. \cite{database118} they correspond to nodes: 85-86-87-88; 34-35-36; 21-22-46; and 52-53-54), while the other nodes are shown as filled circles. The nodes belonging to the four clusters have been arbitrarily selected with the additional requirement of the geographical distance. In this way, the clusters created after control represent groups of nodes, geographically close, that are in symmetry-induced 'islands' with a strong tendency to remain synchronized with each other. In addition, we assume that the original connectedness of the network has to be maintained and, hence, control is performed by applying the CPO method. As result, we have obtained that the target group of symmetries is induced with the addition of 40 links (shown in red in Fig. \ref{fig:PowerGrid}).

\subsection{Central pattern generators}

Central pattern generators (CPGs) are one of the most interesting examples of crossfertilization between biology and neuromorphic engineering. In biology, CPGs are groups of neurons that produce the rhythmic patterns needed for locomotion, possibly processing proprioceptive and heteroceptive information from sensory feedback and commands from higher-level centers \cite{orlovsky1999neuronal}. This notion has become a paradigm for the control of locomotion in bio-inspired robotics, where artificial CPGs are designed for a variety of robot layouts \cite{frasca2004bio}. Artificial CPGs are often modeled as  networks of coupled dynamical units and, remarkably, a class of CPG network architectures, able to reproduce locomotion patterns observed in animals, can be inferred using symmetry methods \cite{collins1993hexapodal,golubitsky1999symmetry}. 
In these works, the networks are given and an analysis of the locomotion gaits that can be generated is carried out. Applying the method illustrated in our work enables the \emph{design} of networks able to generate a given locomotion gait.

\begin{figure}[h]
\begin{center}
{\includegraphics[width=0.24\textwidth]{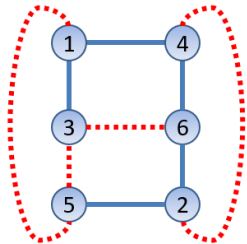}}
\caption{\label{fig:CPG} Design of a CPG network. The original network is depicted with continuous, blue lines. The resulting network also includes the links in dashed, red lines. The controlled network is obtained by applying the CPO method to the original structure with $Q=2$ and $R_1$ and $R_2$ as in (\ref{eq:R1cpg}) and (\ref{eq:R2cpg}).}
\end{center}
\end{figure}

Consider, for instance, a CPG for a six-legged robot. Let us start with the original network depicted in Fig. \ref{fig:CPG} with continuous (blue) lines and suppose (following \cite{collins1993hexapodal}) that the nodes correspond to the following assignment: node 1 - front left leg (L1); node 2 - rear right leg (R3); node 3 - middle left leg (L2); node 4 - front right leg (R1); node 5 - rear left leg (L3); node 6 - middle right leg (R3). Also suppose that the goal is to obtain a CPG able to generate the two gaits denoted as bound-like and alternating metachronical rhythm \cite{collins1993hexapodal}. In the bound-like gait, the legs move in pairs: leg L1 with leg R1, L2 with R2, and L3 with R3. In the alternating metachronical gait, the legs move one at the time in a sequence: L1, R3, L2, R1, L3, R2. Hence, to these gaits, the following permutation matrices can be associated:

\begin{equation}
\label{eq:R1cpg}
\mathrm{R}_1=\left(\begin{array}{llllll}
0 & 0 & 0 & 1 & 0 & 0\\
    0 & 0 & 0 & 0 & 1 & 0\\
    0 & 0 & 0 & 0 & 0 & 1\\
    1 & 0 & 0 & 0 & 0 & 0\\
    0 & 1 & 0 & 0 & 0 & 0\\
    0 & 0 & 1 & 0 & 0 & 0
\end{array}\right)
\end{equation}

\noindent for the bound-like gait, and

\begin{equation}
\label{eq:R2cpg}
\mathrm{R}_2=\left(\begin{array}{llllll}
0 & 1 & 0 & 0 & 0 & 0\\
    0 & 0 & 1 & 0 & 0 & 0\\
    0 & 0 & 0 & 1 & 0 & 0\\
    0 & 0 & 0 & 0 & 1 & 0\\
    0 & 0 & 0 & 0 & 0 & 1\\
    1 & 0 & 0 & 0 & 0 & 0
\end{array}\right)
\end{equation}

\noindent for the alternating metachronal rhytm. Using the CPO method with integer variables as in (\ref{eq:CPOinteger}), the controlled network shown in Fig. \ref{fig:CPG} (including both the original links and the new ones, hallmarked with dashed, red lines) is obtained. Note that the controlled network exactly corresponds to the structure labeled as 'network 3' in \cite{collins1993hexapodal} (Fig. 2), therein shown to generate the above mentioned patterns of locomotion, when periodic oscillators in Birkhoff normal form are used to model the unit dynamics. The method, hence, retrieves the CPG network that in \cite{collins1993hexapodal} was not computed, but provided as a starting point for the analysis of the gaits it can generate via a series of bifurcations occurring before stabilization of the coarsest clustering (that corresponds to GS). 

Although here applied in an exemplificative manner to a case study with few nodes, the approach is applicable to larger CPGs such as the ones considered in \cite{golubitsky1999symmetry}.

\section{Conclusion}

In this work, we have shown that judiciously chosen perturbations to the topology of a given network may enforce an arbitrary set of symmetries and, as a result, may induce desired cluster synchronization dynamics. {Although inducing equitable rather than orbital clusters \cite{schaub2016graph} would be sufficient for the purpose of control, we notice that} the method proposed is efficient as the perturbations to the network structure can be found by only using the generators and not the entire group, which is computationally beneficial for large groups, as the number of generators is significantly smaller than the group order. This type of control is of particular importance in all those circumstances in which the proper functioning of a network requires units coordinated into diverse groups, such as in multi-agent systems performing parallel tasks or in biological systems composed of functionally synchronized clusters. Even in systems normally operating in regime of global synchronization, such as power grids, our approach may be applied to maintain synchronization, in case of failures, in certain parts of the network (a technique known as intentional islanding \cite{balaguer2010control}).

{We have here considered the problem of controlling symmetries and CS in undirected networks of identical oscillators, but our approach can be applied to other frameworks. A first interesting direction for future work is to address the problem in dynamical systems with parameter mismatches among the oscillators, where it has been already shown that approximate CS may emerge \cite{sorrentino2016approximate,wang2019cluster}. It may be also interesting to extend our work to networks with directed links or time-delay couplings, as CS has been shown to be relevant in both scenarios \cite{liu2011cluster,dahms2012cluster}. Finally, the approach described here can be generalized to multilayer networks, where CS may also be observed \cite{della2020symmetries}.}

{Inducing symmetries in a network can also be important for dynamics other than synchronization. A recent study \cite{sorrentino2019symmetries} has in fact pointed out that symmetries play a crucial role in a broad class of network dynamics such as game theory, traffic, and coupled excitable systems. As the first part of our results only concerns the structure of interactions, independently of the specific features of the dynamics taking place at the nodes, it can be purposefully applied to more general frameworks where special features of a dynamical behavior other than synchronization may be desired.}

\end{document}